\documentclass[12pt,nofootinbib,tightenlines,preprintnumbers,amsmath,amssymb]{revtex4}
\date{March 2010}
\usepackage{epsfig}
\usepackage{latexsym}
\usepackage{amssymb}
\usepackage{booktabs}
\usepackage{graphicx}
\newcommand{\be}{\begin{equation}}
\newcommand{\ee}{\end{equation}}
\newcommand{\ba}{\begin{eqnarray}}
\newcommand{\ea}{\end{eqnarray}}
\newcommand{\bi}{\begin{itemize}}
\newcommand{\ei}{\end{itemize}}

\newcommand{\<}{\langle}
\renewcommand{\>}{\rangle}
\newcommand{\eq}{Eq.~}

\newcommand{\fig}{Fig.~}
\newcommand{\tab}{Tab.~}
\newcommand{\la}{\label}

\newcommand{\txts}{\textstyle}

\newcommand{\DPA}{FIGS/}

\renewcommand{\vec}[1]{\boldsymbol{#1}}

\begin{document}
\preprint{MITP/14-041}
\title{Chiral dynamics in the low-temperature phase of QCD}

\author{Bastian B.\ Brandt}
\affiliation{Institut f\"ur theoretische Physik, Universit\"at Regensburg, D-93040 Regensburg}

\author{Anthony Francis, Harvey B.\ Meyer, Daniel Robaina}

\affiliation{\vspace{0.2cm}PRISMA Cluster of Excellence,
Institut f\"ur Kernphysik and Helmholtz~Institut~Mainz,
Johannes Gutenberg-Universit\"at Mainz,
D-55099 Mainz, Germany\vspace{0.2cm}}

\date{\today}

\begin{abstract}
We investigate the low-temperature phase of QCD and the crossover region 
with two light flavors of quarks.
The chiral expansion around the point $(T,m=0)$ in the temperature vs.\  quark-mass  plane
indicates that a sharp real-time excitation exists with the quantum numbers of the pion.
An exact sum rule is derived for the thermal modification of the spectral function 
associated with the axial charge density; the (dominant) pion pole contribution obeys the sum rule.
We determine the two parameters of the pion dispersion relation using lattice QCD simulations
and test the applicability of the chiral expansion.
The time-dependent correlators are also analyzed using the Maximum Entropy Method, yielding consistent results.
Finally, we test the predictions of the chiral expansion around the point $(T=0,m=0)$ 
for the temperature dependence of static observables.
\end{abstract}

\maketitle

\section{Introduction \& main results}

Quark matter at temperatures say $T\gtrsim 30\,{\rm MeV}$ is both of
intrinsic interest as a strongly interacting, quantum relativistic
system, and of relevance in the first few microseconds of the early
universe; see for instance \cite{Brambilla:2014aaa}, chapter D.  It is
studied intensively in heavy-ion collisions.  For small values of the
average up and down quark mass $m$, the system undergoes a transition
from a low-temperature phase where the longest static correlation
length, $m_\pi^{-1}$, scales as $1/\sqrt{m}$ to a high-temperature
phase where the correlation length is largely insensitive to $m$.

One picture of the low-temperature phase that has had significant
phenomenological success is the hadron resonance gas (HRG) model. It
assumes that the thermodynamic properties of the system, including the
conserved charge fluctuations, are given by the sum of the partial
contributions of non-interacting hadron species. The sum extends over
all resonances of mass up to about 2.5GeV, since for most of them the
width is not large compared to the temperature.  The model gives an
economic description of particle yields in heavy-ion collisions (see
the recent \cite{BraunMunzinger:2011ta}, \cite{Stachel:2013zma} and
references therein) and gives a good estimate of the pressure and
charge fluctuations determined in lattice
calculations~\cite{Bazavov:2012jq,Borsanyi:2013bia,Borsanyi:2011sw}.
On the other hand, relatively little is known with certainty about the
spectral functions of local operators (say, the conserved vector
current, the axial current or the energy-momentum tensor) at finite
temperature, which encode the real-time excitations of the
system~\cite{Meyer:2011gj}. The success of the HRG model for static
quantities does not imply that the real-time excitations of the system
are in any sense similar to the ordinary QCD resonances observed at
$T=0$.

A good starting point to investigate the excitations of the thermal medium
is to study what becomes of the pion~\cite{Shuryak:1990ie,Goity:1989gs}. At sufficiently low temperatures
$T\ll T_c$, correlation functions can be computed using chiral
perturbation theory.  The result is that a well-defined pion
quasiparticle persists, with small modifications to the real part of
the pole, and a parametrically small imaginary part~\cite{Schenk:1991xe,Schenk:1993ru,Toublan:1997rr}.
It is not clear how far up in temperature this
treatment can be justified, since the partition function is certainly no longer
dominated by the pions for $T\gtrsim 100\,{\rm MeV}$.  However, the pion is
special in that the Goldstone theorem guarantees the presence of a
divergent static correlation length when $m\to0$ for all
temperatures in the chirally broken phase~\cite{Pisarski:1996mt}.  
If we consider the temperature vs.\  quark-mass  plane
$(T,m)$, this observation suggests an expansion in the quark mass
around the point $(T,0)$. In this case, one gives up on relating the
chiral condensate $\<\bar\psi\psi\>$ and the screening pion amplitude
$f_\pi$ to their $T=0$ counterparts, however the range of
applicability is significantly extended; see \fig\ref{fig:sketch}.  
This is the approach adopted
by Son and Stephanov~\cite{Son:2001ff,Son:2002ci}.  The result of
their analysis is that a pion quasiparticle persists, with a
parametrically small imaginary part compared to the real part of the
pole position. The real part of its dispersion relation is, however,
no longer the relation implied by Lorentz invariance, but rather
\be\la{eq:intro_disprel}
\omega_{\vec k}^2 = u^2(m_\pi^2 + \vec k^2) + \dots
\ee
Here $m_\pi$ is the inverse static correlation length in the
pseudoscalar channel, and $u$, the `pion velocity', is an a priori
unknown function of temperature which can however be related to static
quantities~\cite{Son:2002ci}. Determining $u(T)$ using lattice QCD
for a few temperatures below $T_c$ is one of the main goals of this
paper. We first rederive \eq(\ref{eq:intro_disprel}), present an improved estimator 
for $u(T)$, and show that the spectral function
$\rho_{_{\rm A}}$ of the axial charge density obeys the following
exact sum rule for all temperatures, quark masses and spatial momenta,
\be\la{eq:srA_intro}
\int_{-\infty}^\infty {d\omega\; \omega} \;\rho_{_{\rm A}}(\omega,\vec k)\Big|^{T}_{0} 
= -m\<\bar\psi\psi\>\Big|^{T}_{0}.
\ee
The respective pion pole contributions (which dominate) 
at zero and at finite temperature satisfy the sum rule.

\begin{figure}
\begin{center}
\includegraphics[scale=.4]{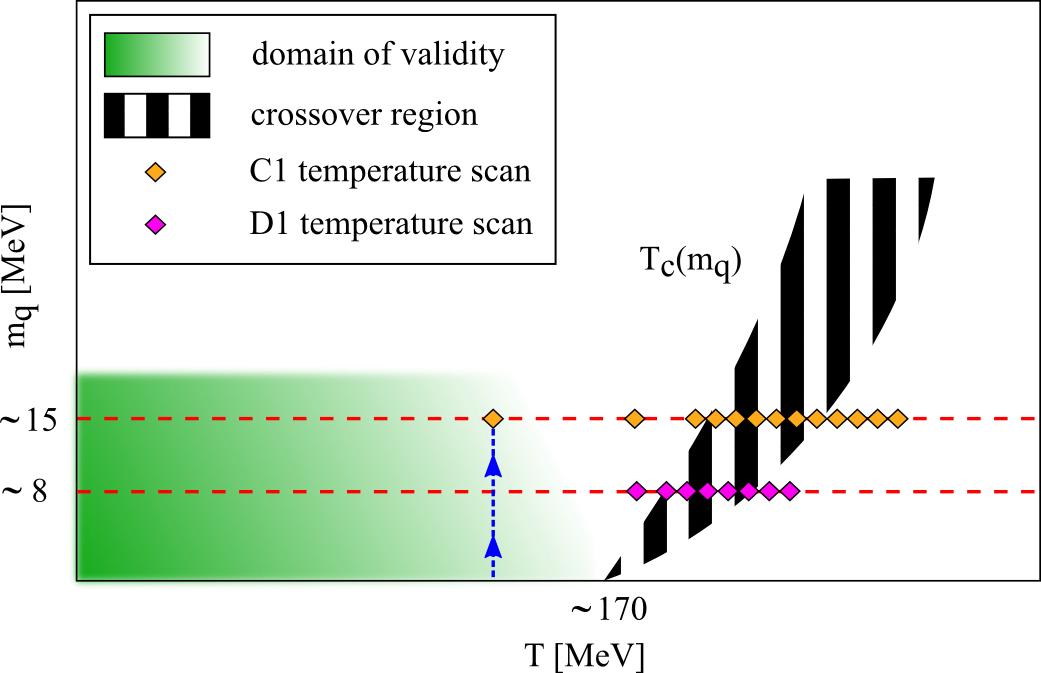}
\end{center}
\caption{Sketch of the domain of validity of the chiral effective field theory 
in the quark mass vs.\ temperature plane. 
The expansion is represented by the blue arrowed vertical line.
The quark mass on the vertical axis is understood to be $\overline{m}^{\overline{\rm MS}}$. 
The value of the critical temperature at the chiral limit $T_{c}(0) \simeq 170$ is taken from  \cite{Brandt:2013mba}.}
\la{fig:sketch}
\end{figure}

We will be working in QCD with two flavors of (O($a$) improved Wilson)
quarks with renormalized masses $8{\rm
  MeV}\leq\overline{m}^{\overline{\rm MS}}\leq 15{\rm MeV}$.  In this
range of quark masses, the transition from the low-temperature to the
high-temperature phase is a crossover, as it is at physical quark
masses. At vanishing $m$, there must be a sharp phase transition,
however its nature is not known with certainty\footnote{The
  possibility considered to be the `standard scenario' is that it is a
  second order phase transition in the 3d O(4) universality class.}.
The crossover temperature, defined conventionally by some observable,
depends quite strongly on the quark mass. The results
of~\cite{Brandt:2013mba} indicate that the pseudocritical temperatures
are $T_{c} = 211(5)\,$MeV and $T_{c} = 193(7)\,$MeV at respectively
$\overline{m}^{\overline{\rm MS}}\simeq 15{\rm MeV}$ and
$\overline{m}^{\overline{\rm MS}}\simeq 8{\rm MeV}$ in the two-flavor
theory.  An extrapolation to $\overline{m}^{\overline{\rm MS}}=0$
yields values between 160MeV and 175MeV for the critical temperature
in the chiral limit~\cite{Brandt:2013mba}. See~\cite{Burger:2011zc,Bornyakov:2009qh}
for other lattice studies of the transition in the two-flavor theory.

The dependence of $T_c$ on the quark mass is sketched in
\fig\ref{fig:sketch}.  The expected domain of applicability of a
chiral expansion around a point $(T,m)$ for $T<T_c(m=0)$ is also
indicated by the shaded region.  We have performed two scans in
temperature at constant renormalized quark mass, indicated by the dots
on the horizontal lines in \fig\ref{fig:sketch}.  Most of the
ensembles considered here thus correspond to the crossover region.
Son and Stephanov have also made predictions for the scaling of the
pion-sector observables~\cite{Son:2001ff} assuming a second-order
phase transition,
\be
f_\pi^2\sim u^2 \sim t^{\nu},\qquad \qquad m_\pi^2 \sim m \,t^{\beta-\nu},
\ee
where $t=(T_c-T)/T_c$ and $\beta,\nu$ are the standard critical exponents.
These scaling predictions are meant to hold as long as $m_\pi\ll m_\sigma \ll
T\simeq T_c$, with $m_\sigma$ the inverse correlation length in the
scalar channel.  We will not be able to test these predictions, 
since it turns out that at the simulated quark masses, the
system still exhibits a very smooth crossover.  We point out, however,
that all observables considered here are well defined for any temperature
and any quark mass; this is in particular true for the estimators of
the quantity $u(T)$ introduced above.  It is the interpretation of the
quantity $u(T)$ as the velocity of a quasiparticle that is uncertain.

The relatively strong dependence of the pseudocritical temperature on
the quark mass tends to reduce the domain of applicability of the
chiral expansion at fixed $T$. For instance, we clearly observe that
the scaling $m_\pi^2\propto m$ is violated at $T\simeq 180\,{\rm
  MeV}$.  Instead the screening pion mass increases (sic!) as the quark mass
is reduced from $15\,$MeV to $8\,$MeV. A plausible explanation is that
at the smaller quark mass, the system is already entering the
crossover region, where the chiral expansion breaks down.

We have found it useful to introduce the following `effective chiral condensate' 
based on the Gell-Mann--Oakes--Renner (GOR) relation,
\be\la{eq:IntroCondGOR}
\left<\bar {\psi} \psi \right>_{\rm GOR} \equiv  -\frac{f^2_\pi m^2_\pi}{m} .
\ee
By construction it has the property that
it tends to the actual chiral condensate when $m\to 0$; it is of order $m$
above $T_c$, and thereby an order parameter with respect to chiral symmetry.
We remark that none of the observables considered here requires the use
of a lattice action preserving chiral symmetry.

The goals of the lattice calculation presented here are the following:
\begin{enumerate}
\item test the validity of the chiral expansion around $(T,m=0)$ and 
compute the pion quasiparticle velocity $u(T)$; 
\item test the chiral expansion around $(T=0,m=0)$;
\item investigate the behavior of $m_\pi$, $f_\pi$ 
and other quantities around the crossover,
where no obvious expansion applies.
\end{enumerate}

\noindent The corresponding results are the following:
\medskip

\noindent {1.} The result for two estimators of the pion velocity is
displayed in \fig\ref{fig:u}.  The reasonable agreement of the two
estimators observed up to $T\simeq 190\,{\rm MeV}$ is a successful
test of the validity of the chiral expansion. It therefore appears
likely that the estimator $u_f(T\simeq 150{\rm MeV})=0.88(2)$ for
$\overline{m}^{\overline{\rm MS}}=15\,{\rm MeV}$ does indeed provide a
valid estimate of the pion quasiparticle velocity. The value indicates
that there is a significant departure from unity, corresponding to a
violation of boost invariance through the presence of the thermal
medium. It shows that, although the hadron resonance gas model
prediction for the thermodynamic potential $(e-3p)/T^4$
\cite{Borsanyi:2013bia} and the charge fluctuations agree well
with lattice results \cite{Borsanyi:2011sw,Bazavov:2012jq}, the
properties of the in-medium excitations can be shifted appreciably
from their $T=0$ counterparts.

The ability to extract the dispersion relation from Euclidean
quantities rests on the do\-mi\-nance of the pion quasiparticle
contribution in the axial charge correlator and in the pseudoscalar
density correlator.  In an attempt to test this dominance explicitly,
we performed a reconstruction of the spectral function based on the
Maximum Entropy Method (MEM). Having investigated the dependence of
the result on the default model which is input to the method, we
conclude that we cannot demonstrate the presence of a peak structure
corresponding to the pion quasiparticle in the spectral
function. However, quite model-independently the spectral weight is
concentrated within $0\leq \omega\lesssim 2.5T$.  The spectral weight
integrated over this interval is correspondingly robust and agrees
with the quantity $(f_\pi/u_f)^2$, which it should if the pion
quasiparticle indeed dominates the correlator.
\medskip

\noindent {2.} Concerning the second goal, we find that at $T\simeq
150\,$MeV, the static screening pion mass and the associated decay constant
$f_\pi$ have changed only by about $5\%$ from their $T=0$ values.
Also the mass of the pion quasiparticle turns out to be 
very close to the $T=0$ pion mass.
These observations are in agreement with the predictions 
of the chiral expansion around 
the point $(m=0,T=0)$~\cite{Gasser:1986vb}.
Only a little higher up in temperature, the decay
constant $f_\pi$ and the correlation length $m_\pi^{-1}$ fall off
rapidly, a behavior no longer described by the chiral expansion.
\medskip

\noindent {3.} \fig\ref{fig:c2} shows the behavior of the effective
condensate defined via the GOR relation. This quantity, in spite of
being a chiral order parameter, varies remarkably slowly throughout
the crossover region. \fig\ref{fig:mpifpi_mq} displays our results
for $m_\pi/T$ and $f_\pi/T$ as a function of temperature for two
different quark masses; the temperature has been rescaled in units of
the quark-mass dependent crossover temperature.  Within the accuracy
of the data, hardly any quark mass dependence is observed.  These
observations indicate that we are still deep in the crossover region and
far from the chiral regime, where one expects a rapid fall-off of the
condensate when $T\stackrel{<}{\longrightarrow}T_c$ and an abrupt rise of $m_\pi$
just above $T_c$.
\medskip

The paper is structured as follows. In section \ref{sec:WI}, we
rederive relation (\ref{eq:intro_disprel}) by exploiting chiral Ward
identities between Euclidean QCD correlation functions. In the
process, we also derive the exact spectral sum rule
(\ref{eq:srA_intro}).  Section \ref{sec:lat} contains the
description of the lattice data and the extraction of the pion
quasiparticle velocity, as well as the comparison with chiral
predictions. Section \ref{sec:mem} presents a study of the
Euclidean-time dependent correlators using the MEM method. Finally, we
give an outlook of how this investigation could be fruitfully
extended.

\section{Chiral Ward identities in the thermal field theory\la{sec:WI}}

We consider Euclideanized QCD with two flavors of degenerate quarks on the space
$S^1\times \mathbb{R}^3$, with the Matsubara cycle $S^1$ of length
$\beta\equiv 1/T$. We label the Euclidean time direction as `0', while the
direction 1, 2 and 3 are of infinite extent; we write $x_\perp \equiv (x_1,x_2)$. 
Unexplained notation follows~\cite{Luscher:1996jn}. 
The Dirac field is a flavor doublet, for instance $\bar\psi(x) = (\bar u(x)~\bar d(x))$.

We define the vector current, axial current and the pseudoscalar density as
\be\la{eq:VAPdef}
V_\mu^a(x) = \bar\psi  \gamma_\mu \frac{\tau^a}{2} \psi(x),
\qquad 
A_\mu^a(x) = \bar\psi  \gamma_\mu \gamma_5 \frac{\tau^a}{2} \psi(x),
\qquad P^a(x)= \bar\psi(x) \gamma_5 \frac{\tau^a}{2} \psi(x).
\ee
where $a\in\{1,2,3\}$ is an adjoint $SU(2)_{\rm isospin}$
 index and $\tau^a$ is a Pauli matrix. The PCAC (partially conserved axial current) 
relation reads
\be\la{eq:PCAC}
\partial_\mu A_\mu^a(x)  = 2m P^a(x),
\ee
where $m$ is the common mass of the up and down quark. \eq(\ref{eq:PCAC}) is valid in any 
on-shell correlation function.
The Ward identities for two-point functions (valid for all $x$; see appendix \ref{sec:apdxWI}) 
that follow from the partial conservation of the axial current are
\be\la{eq:WIA}
\<A^a_\nu(0) \partial_\mu A^b_\mu(x)\> = 2m \< A^a_\nu(0) P^b(x)\>
\ee
where we assume zero isospin chemical potential ($\<V^a_\mu\>=0$ $\forall a,\mu$), and 
\be\la{eq:WIP}
\<P^a(0) \partial_\mu A^b_\mu(x) \> = - \frac{\delta^{ab}}{2} \<\bar\psi\psi\> \delta^{(4)}(x) + 2m\<P^a(0)P^b(x)\>.
\ee

\subsection{Correlators in the massless theory}

Space-time symmetries imply the following form for the $\<P \,\vec A\>$ correlator,
\be\la{eq:PAform}
\int dx_0\; \< P^a(0) \vec A^b(x) \> = \delta^{ab}g(r) \vec e_r, \qquad r= |\vec x|, \quad \vec e_r = \frac{\vec x}{r}.
\ee
Integrating \eq(\ref{eq:WIP}) over $\int_0^\beta dx_0\int_{|\vec x|<R}d^3x$, using the  
 form (\ref{eq:PAform}) and Gauss's theorem, we get
\be\la{eq:gdetd}
g(r) = - \frac{\<\bar\psi\psi\>}{8\pi r^2}.
\ee
This static correlator is thus fully determined by the chiral WI. Integrating \eq(\ref{eq:PAform}) over an 
$x_3=\,$constant plane, 
one obtains\footnote{This last equation can also be obtained directly
 by integrating (\ref{eq:WIP}) over a `slab' $\{x| |x_3| < y_3 \}$ for some positive $y_3$}
\be\la{eq:PAslab}
 \int dx_0\,d^2x_{\perp}\; \<A_3^a(x) P^b(0)\> = -\frac{\delta^{ab}}{4}\, {\rm sign}(x_3)\; \<\bar\psi\psi\> , \qquad m=0.
\ee

A second correlator can also be determined exactly in the massless theory.
Indeed, for $x_0\neq 0$ we have 
\be
\partial_0\int_{r<R} d^3x \< P^a(0)  A_0^b(x) \> =  - \int_{S_R} d\vec\sigma\cdot \<P^a(0)\vec A^b(x)\>.
\ee
We assume that, when  $\<P^a(0)\vec A^a(x)\>$ is expanded in a Fourier series 
in $x_0$, the non-constant modes fall off faster than $1/r^2$.
If we then take the limit $R\to\infty$,  using \eq(\ref{eq:PAform}--\ref{eq:gdetd}) we obtain
\be
\partial_0\int  d^3x\; \< P^a(0) A_0^b(x) \> = \delta^{ab} \frac{\<\bar\psi\psi\>}{2\beta} .
\ee
Thus, since $\<P^a(0) A_0^a(x)\>$ is odd in $x_0$ and in particular vanishes at $x_0=\beta/2$,
\be\la{eq:PA0ref}
\int d^3x\;\<P^a(0)\; A_0^b(x)\> = \delta^{ab}\frac{\<\bar\psi\psi\>}{2\beta}  \Big(x_0-\frac{\beta}{2}\Big).
\ee

\subsection{Correlators at small quark mass: the pion decay constant and the GOR relation}

The power law found in Eqs.\ (\ref{eq:PAform}--\ref{eq:gdetd}) shows
that $P$ couples to a massless screening particle.  The main idea in
the following is to obtain the residue of the poles in the chiral
limit, where they are determined by chiral Ward identities, and to use
those at small but finite quark mass.

At finite quark mass, we expect\footnote{This equation defines $m_\pi$.}
\be\la{eq:APgen}
 \int dx_0\,d^2x_{\perp}\; \<A_3^a(x) P^b(0)\> \stackrel{|x_3|\to\infty}{=}  \delta^{ab}{\rm sign}(x_3) \,c(m)\, \exp(-m_\pi |x_3|),
\ee
with $c(0)= -\frac{1}{4} \<\bar\psi\psi\>$ in view of \eq(\ref{eq:PAslab}). Since the PCAC relation (\ref{eq:PCAC}) implies 
\be
\partial_3 \int dx_0\,d^2x_{\perp}\; \<A_3^a(x) P^b(0)\> =  2m \int dx_0\,d^2x_{\perp}\; \<P^a(0)\,  P^b(x)\>, 
\ee
we learn from (\ref{eq:APgen}) that close to the chiral limit,
\be
\int dx_0\,d^2x_{\perp}\;\<P^a(0)\,  P^b(x)\> =  \delta^{ab}\frac{\<\bar\psi\psi\>\, m_\pi}{8m} \exp(-m_\pi |x_3|).
\ee
This equation shows that the correlation function of the pseudoscalar density admits a pole at $m_\pi$ 
with residue $\frac{m_\pi^2 \<\bar\psi\psi\>}{4m}$. Consequently, since the scalar propagator is $\frac{\exp(-m_\pi r)}{4\pi r}$
in three dimensions, we can write 
\be\la{eq:PPstat}
 \int dx_0 \;\<P^a(0) \;P^b(x)\> \stackrel{r\to\infty}{=}  \delta^{ab} \frac{m_\pi^2 \<\bar\psi\psi\>}{4m}\; \frac{\exp(-m_\pi r)}{4\pi r}.
\ee
Now returning to \eq(\ref{eq:APgen}), multiplying both sides by $2m$ and using the PCAC relation shows that 
close to the chiral limit,
\be\la{eq:A3A3ch}
\int dx_0\,d^2x_{\perp}\; \<A_3^a(x) A_3^b(0)\> \stackrel{|x_3|\to\infty}{=} -  \delta^{ab}\frac{m\<\bar\psi\psi\>}{2m_\pi} \exp(-m_\pi |x_3|).
\ee
We know that the correlator $\int dx_0 \<P^a(0)\vec A^a(x)\>$ 
is non-zero at $m=0$; therefore the coupling of $P$ to the Goldstone boson cannot vanish at $m=0$ --- 
consistently with Goldstone's theorem.
Since the residue at the pion pole in the correlator (\ref{eq:PPstat}) cannot diverge in the chiral limit, 
we conclude that  $m_\pi^2\sim  m$. 

The scaling of $m_\pi$ with the quark mass motivates the definition of $f_\pi$ (for any value of the quark mass) via
\be\la{eq:A3A3}
\int dx_0\,d^2x_{\perp} \<A_3^a(x) A_3^b(0)\> = \frac{ \delta^{ab}}{2} f_\pi^2 m_\pi e^{-m_\pi |x_3|},\qquad  |x_3|\to\infty.
\ee
Comparison with the chiral prediction (\ref{eq:A3A3ch}) shows that 
\be\la{eq:GOR}
f_\pi^2 m_\pi^2 = -m\< \bar\psi\psi\>,\qquad m\to 0,
\ee
in particular $f_\pi$ has a finite, non-vanishing limit when $m\to 0$ as long as $\< \bar\psi\psi\>$ is finite.

\subsection{Spectral functions \la{sec:spf}}

Relation (\ref{eq:PA0ref}) shows that the pseudoscalar density and the
axial charge density couple to a (real-time) massless excitation in
the chiral limit.  The goal is now to compute the dispersion relation
of this excitation for small quark masses and spatial momenta.

We recall the relation between the spectral function and the Euclidean correlator
for the following cases,
\ba\la{eq:SF1}
\delta^{ab}G_{\rm P}(x_0,\vec k) \equiv \int d^3x \; e^{-i\vec k\cdot\vec x}\; \<P^a(0) P^b(x)\> 
&=&  \delta^{ab}\int_0^\infty d\omega\, \rho_{_{\rm P}}(\omega,k)\, \frac{\cosh(\omega(\beta/2-x_0))}{\sinh(\omega\beta/2)}\,,\qquad 
\\
\la{eq:SF2}
 \delta^{ab}G_{\rm AP}(x_0,\vec k) \equiv  \int d^3x \; e^{-i\vec k\cdot\vec x}\; \<P^a(0) A_0^b(x)\> 
&=&  \delta^{ab} \int_0^\infty d\omega\, \rho_{_{\rm AP}}(\omega,k)\, \frac{\sinh(\omega(\beta/2-x_0))}{\sinh(\omega\beta/2)}\,,\qquad
\\
\la{eq:SFA}
 \delta^{ab} G_{\rm A}(x_0,\vec k) \equiv \int d^3x\; e^{-i\vec k\cdot\vec x}\;\< A_0^a(0) A_0^b(x)\> 
&=&  \delta^{ab} \int_0^\infty d\omega\, \rho_{_{\rm A}}(\omega,k)\; 
\,\frac{\cosh(\omega(\beta/2-x_0))}{\sinh(\omega\beta/2)}\,.\qquad 
\ea
The PCAC relation (\ref{eq:PCAC}) implies
\ba\la{eq:PCACrhoP}
2m\,\rho_{_{\rm P}}(\omega,0) &=& - \omega\, \rho_{_{\rm AP}}(\omega,0),
\\
\omega\, \rho_{_{\rm A}}(\omega,0)  &=&  2m\, \rho_{_{\rm AP}}(\omega,0).
\la{eq:PCACrhoA}
\ea
Equation (\ref{eq:PA0ref}), which is an exact expression in the chiral limit, shows that 
\be
\rho_{_{\rm AP}}(\omega,0) = -\frac{\<\bar\psi\psi\>}{2} \delta(\omega)\qquad \quad (m=0).
\ee
Since $P(x)$ and $A_0(x)$ couple to a  massless excitation at $m=k=0$, they must also couple 
to an excitation when $m$ and $k$ are small but finite. In the following we assume that 
the imaginary part of the pole is negligible compared to its real part. An analysis in the 
hydrodynamic framework supports this assumption~\cite{Son:2002ci}, as well as the chiral
expansion around $T=0$~\cite{Schenk:1993ru}. We thus write the ansatz
\be\la{eq:rhoP}
\rho_{_{\rm P}}(\omega,k) = {\rm sign}(\omega) C(k^2)  \delta(\omega^2-\omega_{\vec k}^2) + \dots
\ee
for the spectral function of the pseudoscalar density. We must have $\omega_{\vec k}\to 0$ 
when $m,k\to 0$ and the function $C(k^2)$ 
is the residue of the pole in $\omega^2$ and is non-vanishing when $m_\pi,k\to0$.
We now show that the dispersion relation is of the form
\be\la{eq:disprel}
\omega_{\vec k}^2 = u^2 ( m_\pi^2 + \vec k^2)+ {\rm O}((\vec k^2)^2).
\ee
The key observation  is that we know the static correlator, \eq(\ref{eq:PPstat});
it is proportional to a three-dimensional scalar propagator. The static correlator 
can be expressed in terms of the spectral function as follows (see for instance~\cite{Meyer:2008sn}),
\ba
\int dx_0 \;\<P^a(0) \;P^b(x)\> 
&=&  2\delta^{ab}\lim_{\epsilon\to0} \int \frac{d^3k}{(2\pi)^3} e^{i\vec k\cdot \vec x}
\int_0^\infty \frac{d\omega}{\omega} e^{-\epsilon\omega}\rho_{_{\rm P}}(\omega,k)
\nonumber
\\ &=& \delta^{ab}\int \frac{d^3k}{(2\pi)^3} e^{i\vec k\cdot \vec x} \frac{C(k^2)}{\omega_{\vec k}^2}+\dots
\ea
Comparing with \eq(\ref{eq:PPstat}), we see that $\omega_k^2$ must be
proportional to $m_\pi^2+\vec k^2$. Calling the proportionality factor $u^2$,
we have  proved \eq(\ref{eq:disprel}) and we then have
\be\la{eq:CP}
C(k^2) = -\frac{\<\bar\psi\psi\>^2\,u^2}{4 f_\pi^2}
\ee
in the limit of small $m_\pi$ and $k$.
Relations (\ref{eq:PCACrhoP}--\ref{eq:PCACrhoA}) now lead to 
\ba
\la{eq:rhoAP}
\rho_{_{\rm AP}}(\omega,0) &=& -\frac{\omega_{\vec 0} \<\bar\psi\psi\>}{2}\, \delta(\omega^2-\omega_{\vec 0}^2) + \dots,
\\
\la{eq:rhoA}
\rho_{_{\rm A}}(\omega,0) &=& {\rm sign}(\omega) f_\pi^2 m_\pi^2 \,\delta(\omega^2-\omega_{\vec 0}^2) + \dots
\ea

\subsection{An exact sum rule for $\rho_{_{\rm A}}(\omega,q)$ \la{sec:sr}}

In appendix \ref{sec:apdxWI}, we show that the chiral Ward identities, together with the ultraviolet properties
of the axial current correlator, imply the following exact sum rule for the axial current spectral function 
\be\la{sr:A}
\int_{-\infty}^\infty {d\omega\; \omega} \;\rho_{_{\rm A}}(\omega,\vec k)\Big|^{T}_{0} = -m\<\bar\psi\psi\>\Big|^{T}_{0}.
\ee
This equation is valid at vanishing chemical potential, but for any quark mass;
it is to be compared to the corresponding sum rule 
in the vector channel ($\<V_0 V_0\>$, \cite{Bernecker:2011gh,Brandt:2012jc}),
\be\la{sr:V}
\int_{-\infty}^\infty {d\omega\; \omega} \;\rho_{_{\rm V}}(\omega,\vec k)\Big|^{T}_{0} = 0.
\ee
The symbol $\{\dots\}\Big|^{T}_{0}$ means that the zero-temperature
contribution is subtracted. The subtraction is necessary to make the
integral over frequency convergent. One easily checks that the
pion-quasiparticle contribution to $\rho_{_{\rm A}}(\omega,0)$ given
in (\ref{eq:rhoA}) and the $T=0$ pion contribution satisfy the sum
rule (\ref{sr:A}).

The sum rules (\ref{sr:A}) and (\ref{sr:V}) are complementary to the
sum rules derived in \cite{Kapusta:1993hq} in the massless theory.
For $m=0$, \eq(\ref{sr:A}--\ref{sr:V}) are consistent with the sum
rule `II-L' given in \cite{Kapusta:1993hq} upon substracting the $T=0$
contributions.

\subsection{Expressing $u^2$ in terms of static quantities}

The parameter $u$ can be obtained from $G_{\rm A}(x_0,\vec 0)$, at sufficiently small quark mass, by noting that
\be\la{eq:w0simple}
\omega_{\vec 0}^2 = \frac{\partial_0^2 G_{\rm A}(x_0,\vec 0)}{G_{\rm A}(x_0,\vec 0)}\Big|_{x_0=\beta/2} = 
-4m^2 \frac{G_{\rm P}(x_0,\vec 0)}{G_{\rm A}(x_0,\vec 0)}\Big|_{x_0=\beta/2}
\ee

The chiral Ward identities allow one to express $\partial_0^2 G_{\rm A}(x_0,\vec 0)$ in terms of 
$f_\pi$, $m_\pi$ and $\omega_{\vec 0}$.
Using the spectral function (\ref{eq:rhoA}), one obtains
\be\la{eq:Goopp}
\partial_0^2 G_{\rm A}(x_0,\vec 0) = \frac{f_\pi^2 m_\pi^2 \omega_{\vec 0}}{2}
\;\frac{\cosh(\omega_{\vec 0}(\beta/2-x_0))}{\sinh(\omega_{\vec 0}\beta/2)}.
\ee
Inserting expression (\ref{eq:Goopp}) into \eq(\ref{eq:w0simple}) yields the following algebraic equation for $u$,
\be\la{eq:ush}
u\,\sinh(u\, m_\pi \beta/2) = \frac{f_\pi^2 m_\pi}{2 G_{\rm A}(\beta/2,\vec 0)}.
\ee
This equation provides a way to extract the velocity $u$ from Euclidean correlation functions.
It is valid throughout the the shaded region in \fig\ref{fig:sketch}, i.e.\ for sufficiently 
small quark masses and for all $T<T_c(m=0)$.
In the massless case, this relation is equivalent to the result of Son and Stephanov~\cite{Son:2002ci},
\be\la{eq:uSS}
u^2 = \frac{f_\pi^2}{\int_0^\beta dx_0 \;G_{\rm A}(x_0,\vec 0)} \qquad \qquad (m=0). 
\ee
The axial susceptibility appearing in the denominator of (\ref{eq:uSS}) 
however contains an ultraviolet divergence at any non-vanishing quark mass.
It is therefore not practical to use in lattice calculations.

\section{Lattice calculation\la{sec:lat}}

In this section we describe the numerical lattice QCD calculation of
the temperature dependent parameters $u$ and $m_\pi$ that characterize the
pion dispersion relation; see \eq(\ref{eq:disprel}). All finite
temperature correlation functions are measured on a set of dynamical
gauge ensembles with two mass degenerate quark flavors covering a
temperature range $150\leq T\leq 235\,{\rm MeV}$. We use the
plaquette gauge action and the O($a$) improved Wilson fermion action
with a non-perturbatively determined $c_{\rm{sw}}$ coefficient
\cite{Jansen:1998mx}. The configurations were generated using the
MP-HMC algorithm \cite{Hasenbusch:2001ne,Hasenbusch:2002ai} following
the implementation described in \cite{Marinkovic:2010eg} based on
L\"uscher's DD-HMC package~\cite{CLScode}.

Two scans in temperature were carried out on lattices of size
$16\times32^3$, where the short direction is interpreted as time and
therefore $T=1/(16a)$ and the spatial extent is $L=32a$.  The gluon
fields have periodic boundary conditions in all directions, while the
quark fields are periodic in space and antiperiodic in time.  The
temperature is varied by varying the bare coupling $g_0^2$, which
amounts to varying the lattice spacing at fixed `aspect ratio' $LT=2$.
The scale setting was done via the Sommer parameter
\cite{Sommer:1993ce}. We use a quadratic interpolation of $\log(r/a)$
based on the data given in \cite{Fritzsch:2012wq} to relate the
lattice spacings at two values of the bare coupling.  The absolute
scale setting, $a/{\rm fm}$, is done using the value $r_0 =
0.503(10)\,$fm~\cite{Fritzsch:2012wq}. 

The two scans correspond to two quark masses of respectively about
8MeV and 15MeV, where the bare quark mass is tuned to keep the
renormalized quark mass constant (see \fig\ref{fig:awi}).  The quark mass
is given in the $\overline{\rm{MS}}$ scheme at a scale $\mu=2$GeV for
which we used the renormalization factors $Z_A(g^2_0)$ and
$Z_P=0.5184(53)$ from \cite{Fritzsch:2012wq} as well as the conversion
factor from the Schr\"odinger Functional (SF) to the $\overline{\rm
  MS}$ scheme, which is 0.968(20) \cite{Fritzsch:2012wq}.

We use the standard definition for the quark mass that comes from the PCAC 
 relation \cite{Bochicchio:1985xa,Luscher:1996ug}
\be
m_{\rm PCAC}(x_3) = \frac{1}{2} \frac{\int dx_0 d^2x_{\perp} 
\left<\partial^{\rm imp}_3 A^{a,{\rm imp}}_3(x) P^a(0)\right>}{\int dx_0 d^2x_{\perp} 
\left<P^b(x)P^b(0)\right>}, \qquad x_\perp = (x_1, x_2),
\ee
where in the improvement process
\be
A_{\mu}^a \longrightarrow A^{a,{\rm imp}}_{\mu} = A^a_{\mu} + ac_A \partial^{\rm imp}_\mu P^a.
\ee
The derivative $\partial^{\rm imp}_\mu$ is the improved lattice
discretized version of the derivative following
\cite{Guagnelli:2000jw}. The non-perturbatively calculated coefficient
$c_A$ was taken from \cite{DellaMorte:2005se}. Notice that since the
PCAC relation is an operator identity, we are free to choose the
direction in which we define the quark mass -- any dependence on the
direction must therefore amount to a discretization error. On our lattices,
the spatial direction is longer, so measuring along these directions
we obtain a longer plateau and thus, smaller errors. The extraction is
carried out by performing a fit to a constant in the range where a
plateau is observed. Within errors the PCAC masses measured in the
time and in spatial directions agree. 

The two scans, called C1 and D1, have respectively pseudocritical
temperatures of $T_{c} = 211(5)\,$MeV and $T_{c} =
193(7)\,$MeV~\cite{Brandt:2013mba}.  For instance, in
\fig\ref{fig:awi}, we observe that in scan C1, our renormalized quark
mass in physical units is approximately constant up to $T=211$MeV
where the phase transition to the deconfined phase is estimated to
occur. This means that the ensembles presented in Tabs.\ \ref{tab:C1}
and \ref{tab:D1} follow to a good approximation `lines of constant
physics' and can be interpreted as temperature scans at fixed quark
mass.

Statistical errors on the observables are calculated using the
jackknife method.  In plots, only the statistical error from our
simulations are displayed; the error from renormalization factors and
the scale setting uncertainty should be added in quadrature to obtain
the full uncertainty.

\begin{figure}[t]
\begin{center}
\includegraphics[width=.5\textwidth]{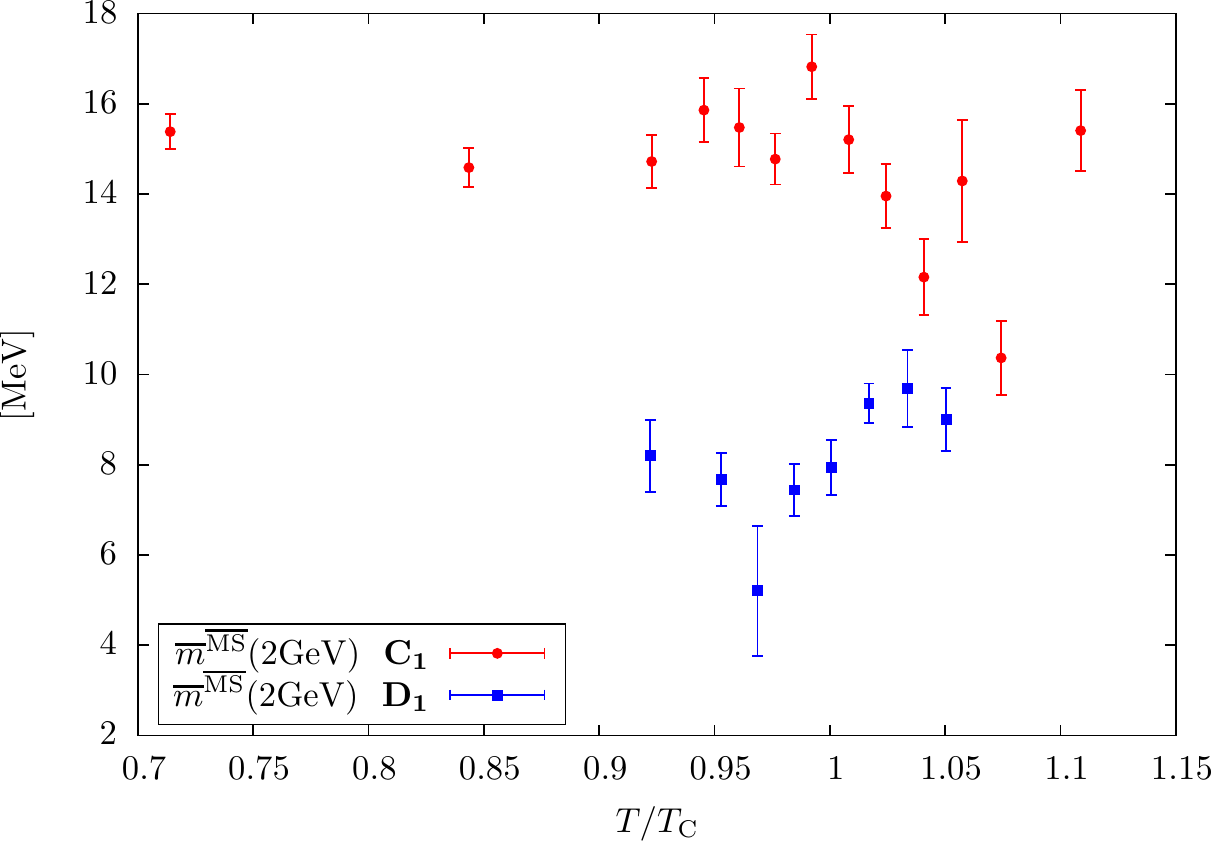}
\caption{Renormalized quark mass in physical units in the $\overline{\rm{MS}}$ scheme 
for both temperature scans (C1 and D1). 
For the C1 scan, $T_{c} = 211(5)\,$MeV, and for D1, $T_{c} = 193(7)\,$MeV.}
\la{fig:awi}
\end{center}
\end{figure}

\subsection{Basic observables}
In this section we describe the calculation of the following observables:
\begin{itemize}
\item the midpoint of the axial charge correlator in the time direction, $G_{{\rm A}}(\beta/2,\vec 0)$;
\item the midpoint of the pseudoscalar correlator  correlator in the time direction, $G_{{\rm P}}(\beta/2,\vec 0)$;
\item the screening pion mass $m_\pi$;
\item the screening pion decay constant $f_\pi$.
\end{itemize}

The values of the correlators $G_{{\rm A}}(\beta/2,\vec 0)$ and $G_{{\rm P}}(\beta/2,\vec 0)$
at $x_0=\beta/2$ are displayed in \fig\ref{fig:GAGP}.  While the former only exhibits a
mild temperature dependence, the latter quantity is strongly temperature dependent.
Since $\partial_0^2 G_{\rm A}(x_0)= -4m^2 G_{\rm P}(x_0)$, this observation means that the axial charge
correlator becomes flatter as a function of $x_0$. It shows that the spectral density $\rho_{\rm _A}(\omega,0)$ 
must concentrate around the origin as the temperature rises.

\subsubsection{Extraction of $m_\pi$}
In order to extract the `screening' pion mass, we compute the symmetrized
pseudoscalar-pseudoscalar screening Euclidean correlator along a spatial direction,
\be
\delta^{ab}\;G^{\rm s}_{\rm P}(x_3) = \int dx_0 d^2x_{\perp} \left<P^a(x)P^b(0)\right> ,\qquad \quad 
G^{\rm s}_{\rm P}(x_3) \stackrel{|x_3| \to \infty}{\sim} e^{-m_\pi |x_3|}
\ee
At long distances, it is dominated by the lowest lying state with pseudoscalar quantum numbers, 
which we call the `screening pion'.
In practice, a two state fit to the correlation function via 
Levenberg-Marquardt's method \cite{press_numerical_1992} is performed using an ansatz of the form
\be
G^{\rm s}_{\rm P}(x_3) = A^P_1 \cosh[m^P_1(x_3-L/2)] + A^P_2 \cosh[m^P_2(x_3-L/2)].
\ee
To initialize the fit-routine we use as input parameter for $m^P_1\doteq m_\pi$ 
an averaged value of the `coshmass' $m_{\rm cosh}(x_3)$ defined as the positive root of the following equation,
\be
\frac{G^{\rm s}_{\rm P}(x_3)}{G^{\rm s}_{\rm P}(x_3+a)} = \frac{\cosh[m_{\rm cosh}(x_3+a/2)(x_3- L/2)]}{\cosh[m_{\rm cosh}(x_3+a/2)(x_3+a-L/2)]}\;;
\ee
In order to be sure that the ground state is isolated
one can repeat the fit to the correlation function for different fit
windows, leaving out points that are furthest away from the middle
point $x_3=L/2$ of the correlator. We choose for the $m_\pi$ result
quoted in table \ref{tab:C1summ} a value corresponding to a small
$\chi^2/{\rm d.o.f}$ which is stable under small variations of the fit window.
The result for $m_\pi$ obtained in this way is close, in value and in its uncertainty,
to  $m_{\rm cosh}$ around $x_3=L/2$; see \fig\ref{fig:coshmass}.

The temperature dependence of $m_\pi$ is illustrated in the left panel of \fig\ref{fig:mfc}.
We observe that the correlation length in the thermal medium becomes shorter as the temperature 
increases, and is about half as long at the crossover as it is at zero-temperature.

\begin{figure}[t]
    \includegraphics[width=.48\textwidth]{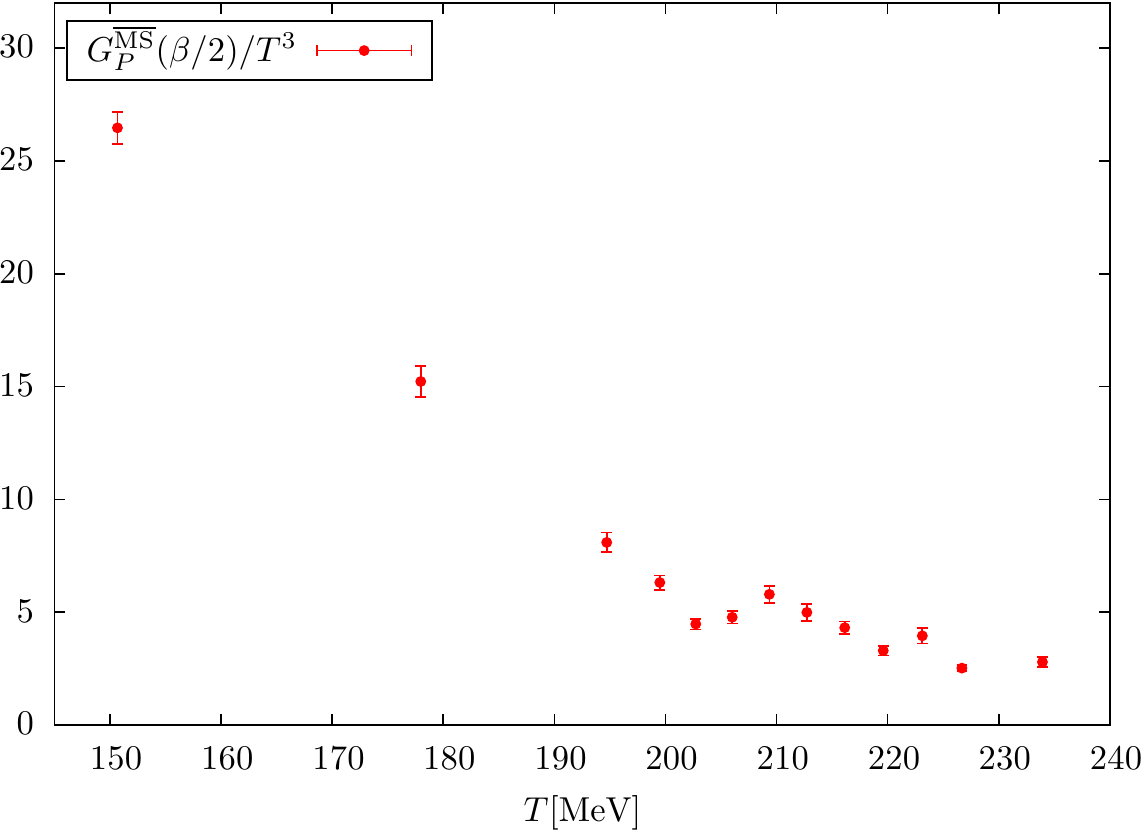}
    \includegraphics[width=.48\textwidth]{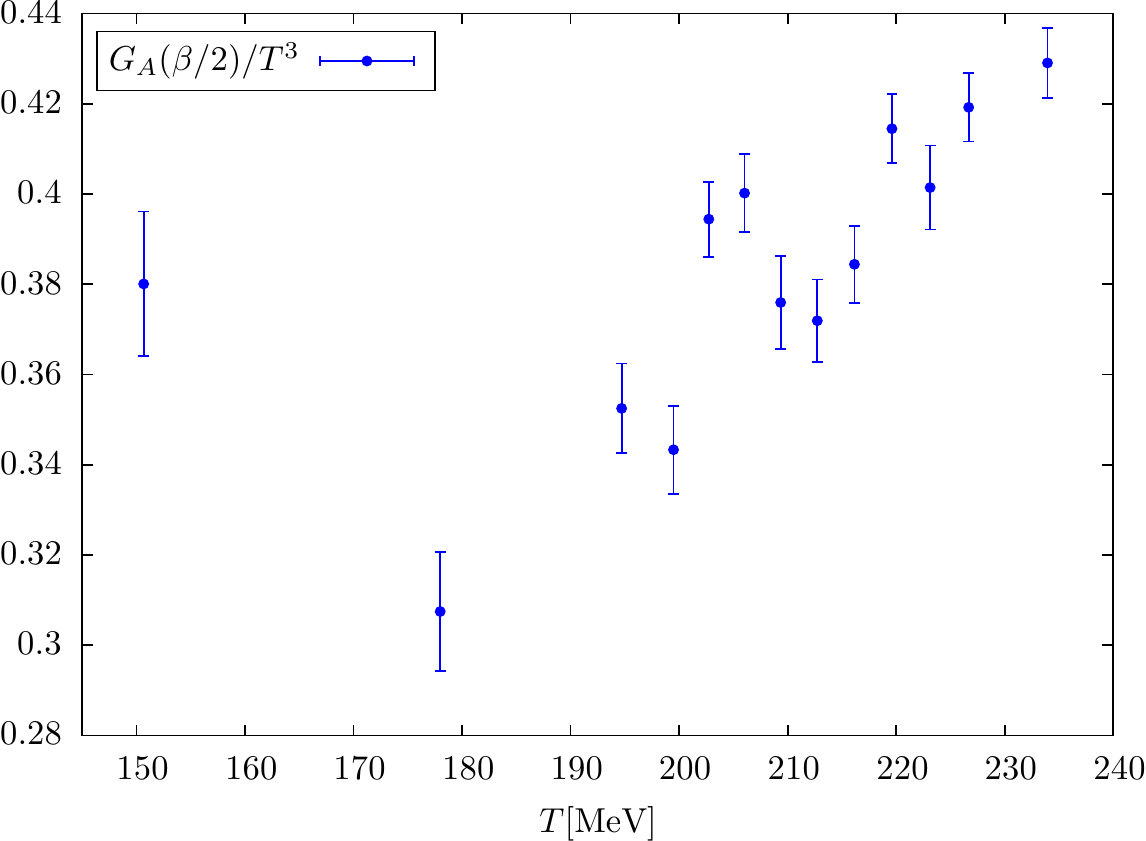}
   \caption{Midpoints of the renormalized correlators $G_{\rm A}(x_0)$ and $G_{\rm P}(x_0)$.
     The former was renormalized via multiplication with $Z_{\rm A}^2$, the latter via multiplication with 
$Z^2_{\rm P}$ as well as the conversion factor from the SF to the $\overline{\rm{MS}}$ scheme at the scale $\mu=2$GeV.
All data from the C1 temperature scan.}
\la{fig:GAGP}
\end{figure}

\subsubsection{Extraction of $f_\pi$}

We extract $f_\pi$ from the correlation function
\be
\delta^{ab}\,G^{\rm s}_A(x_3) = \int dx_0 d^2x_{\perp} \left <A^{a,{\rm imp}}_3(x) A^{b,{\rm imp}}_3(0)\right> 
\stackrel{|x_3| \to \infty}{=} \frac{\delta^{ab}}{2}f^2_\pi m_\pi e^{-m_\pi x_3}
\ee
Because of the noisier behavior of this correlator, it turns out that the fit to this
correlation function is more stable using a 1-state-fit rather than a
2-state-fit. Since $G^{\rm s}_A$ is symmetric around $x_3=L/2$ we use
an ansatz of the form
\be
G^{\rm s}_A(x_3)=A^A_1 \cosh[m^A_1(x_3-L/2)].
\ee
For stability reasons we put $m^A_1 = m_\pi$ by hand since this
quantity is already known from the $G^s_{\rm P}$-fit. This reduces the number
of parameters to one. By repeating the procedure for different fit
windows as explained above, we select the final value for $A^A_1$ by choosing a fit
which has a low $\chi^2/{\rm d.o.f}$. The relation between $A^A_1 $ and $f_\pi$ reads 
\be
f^2_\pi = \frac{2\,A^A_1 \sinh(m_\pi L/2)}{m_\pi}.
\ee
The temperature dependence of $f_\pi$ in the C1 temperature scan is displayed in the right panel
of \fig\ref{fig:mfc}.  We observe a reduction of $f_\pi$ as the temperature increases,
reaching a value of about one third its zero-temperature value around the crossover.

\begin{figure}[t]
\includegraphics[width=.48\textwidth]{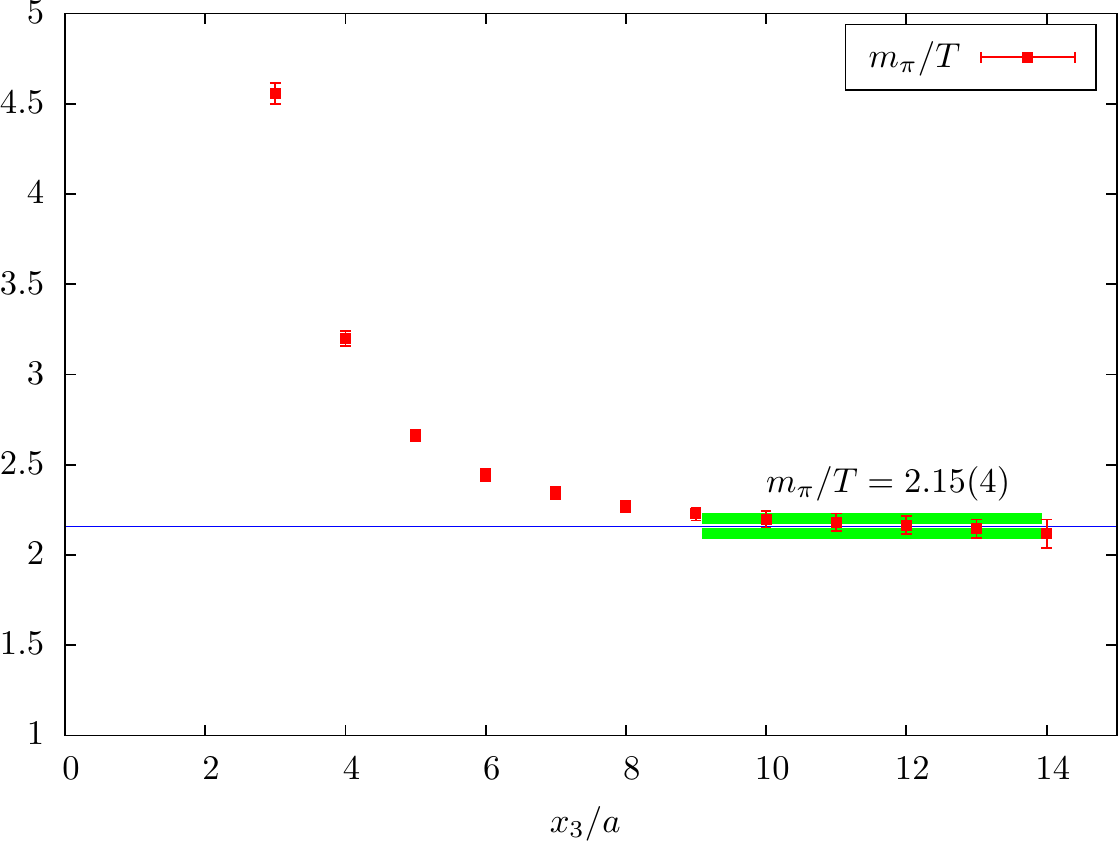}
\caption{Example of an effective-mass plot showing $m_{\rm cosh}(x_3+a/2)$
for the pseudoscalar density two-point function in the $x_3$-direction 
in the C1 scan at $T=150\,{\rm MeV}$. 
The result of the fit to the correlation function is represented by a ($1\sigma$) band.
Here the chosen fit-window was 26, which
corresponds to ignoring the three points closest to each operator,
and the (uncorrelated) $\chi^2/{\rm d.o.f}$ amounts to $0.05$.}
\la{fig:coshmass}
\end{figure}

\begin{figure}[t]
\includegraphics[width=.48\textwidth]{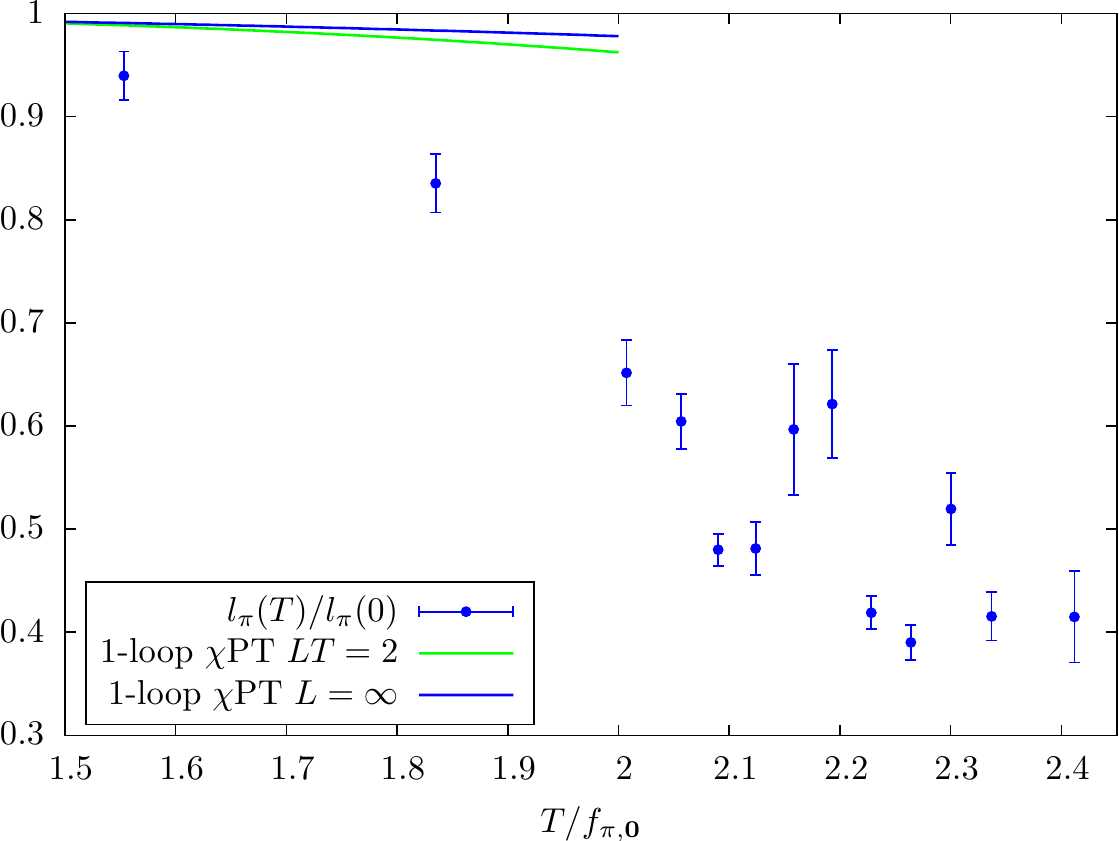}
\includegraphics[width=.48\textwidth]{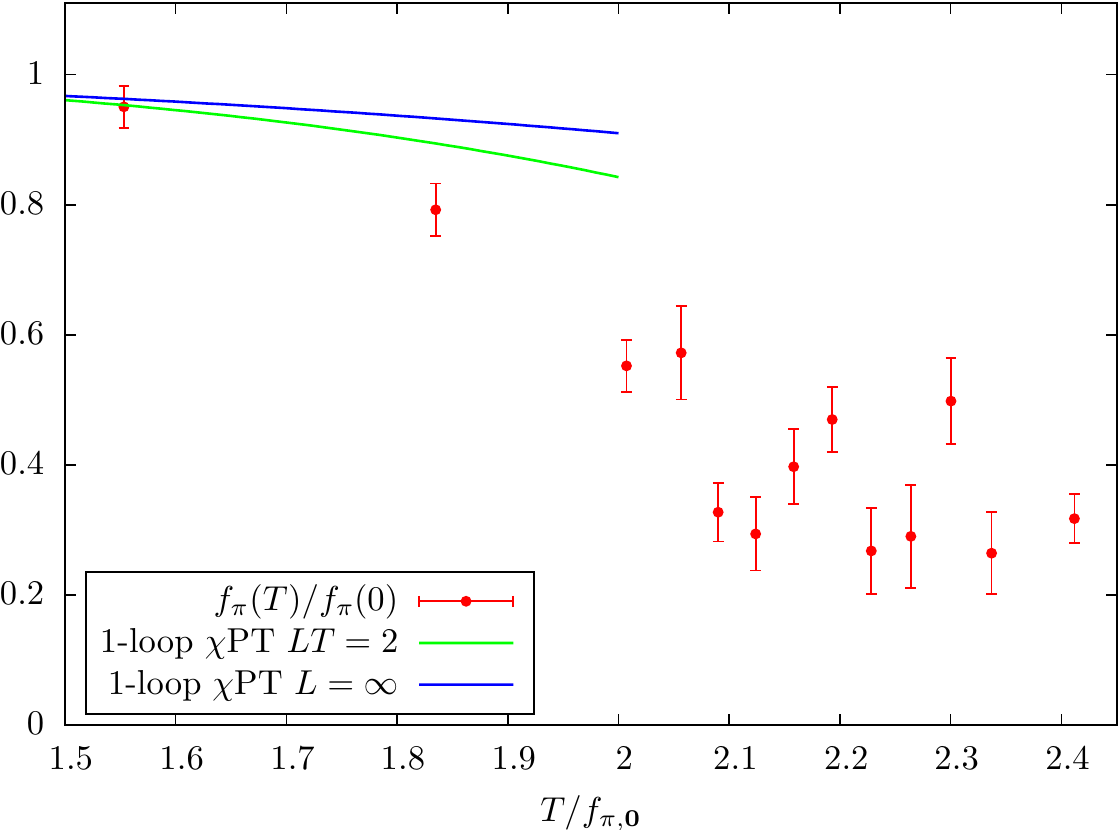}
\caption{Inverse screening mass $l_\pi\equiv m_\pi^{-1}$ (left) and 
screening pion `decay constant' (right) in the C1 scan, divided 
by the same quantity at $T\simeq0$ extracted from the A5 ensemble. 
The displayed error bars represent the statistical errors originating from 
the ensembles of the C1 scan and from ensemble A5.}
\la{fig:mfc}
\end{figure}

\subsubsection{Chiral condensate $\left<\bar{\psi}\psi\right>$}
Using the Gell-Mann--Oakes--Renner relation \cite{GellMann:1968rz}, 
one can define an effective chiral condensate as follows
(see \eq\ref{eq:IntroCondGOR}),
\be\la{eq:CondGOR}
\left<\bar {\psi} \psi \right>^{\overline {\rm MS}}_{\rm GOR} = -\frac{f^2_\pi m^2_\pi}{\overline{m}^{\overline {\rm MS}}} .
\ee
Since $m_\pi \sim T$ and $f_\pi\sim m$ above $T_c$,
$ \left<\bar {\psi} \psi \right>^{\overline {\rm MS}}_{\rm GOR}$ is of order $m$ 
above $T_c$; at high temperatures, it is expected to grow as $mT^2$.

The behavior of the effective chiral condensate is displayed in  \fig\ref{fig:c2}.
We find it to be weakly temperature dependent around $T_c$. 
It illustrates how smooth the crossover is at the quark mass
used in the temperature scan C1: around $T=200\,{\rm MeV}$, 
$ |\left<\bar {\psi} \psi \right>^{_{\overline {\rm MS}}}_{\rm GOR}|^{1/3}$
 only appears to be about $10\%$ lower than at zero temperature.

\begin{figure}[t]
\begin{center}
\includegraphics[width=.6\textwidth]{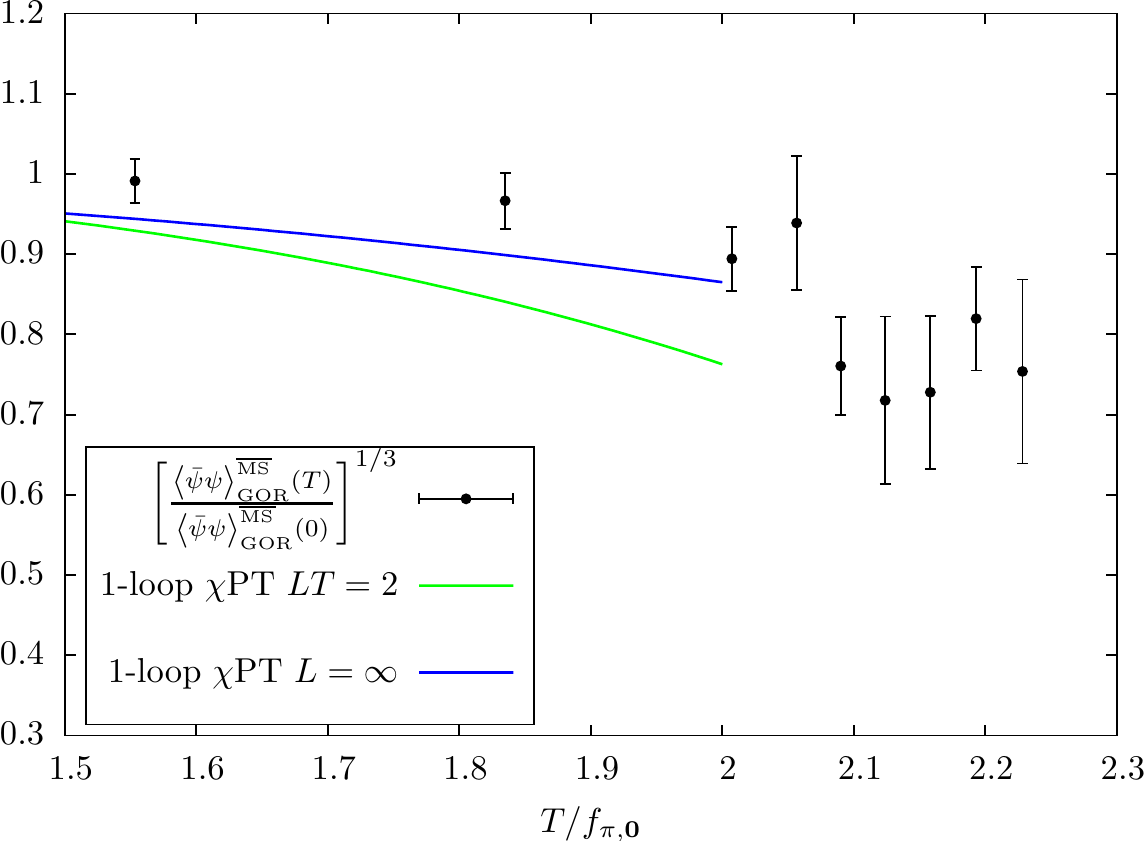}
\end{center}
\caption{
Effective chiral condensate defined from the GOR relation, divided by its $T=0$ counterpart,
in the temperature scan C1.
In addition, the predictions of \cite{Gasser:1986vb} both for 
the infinite volume limit and for our finite lattice volume  are displayed. 
The temperature is given in units of the zero-temperature decay constant  $f_{\pi, \bf{0}}$. }
\la{fig:c2}
\end{figure}

\subsection{Lattice estimators for the pion velocity}

\begin{figure}[t]
\begin{center}
    \includegraphics[width=.48\textwidth]{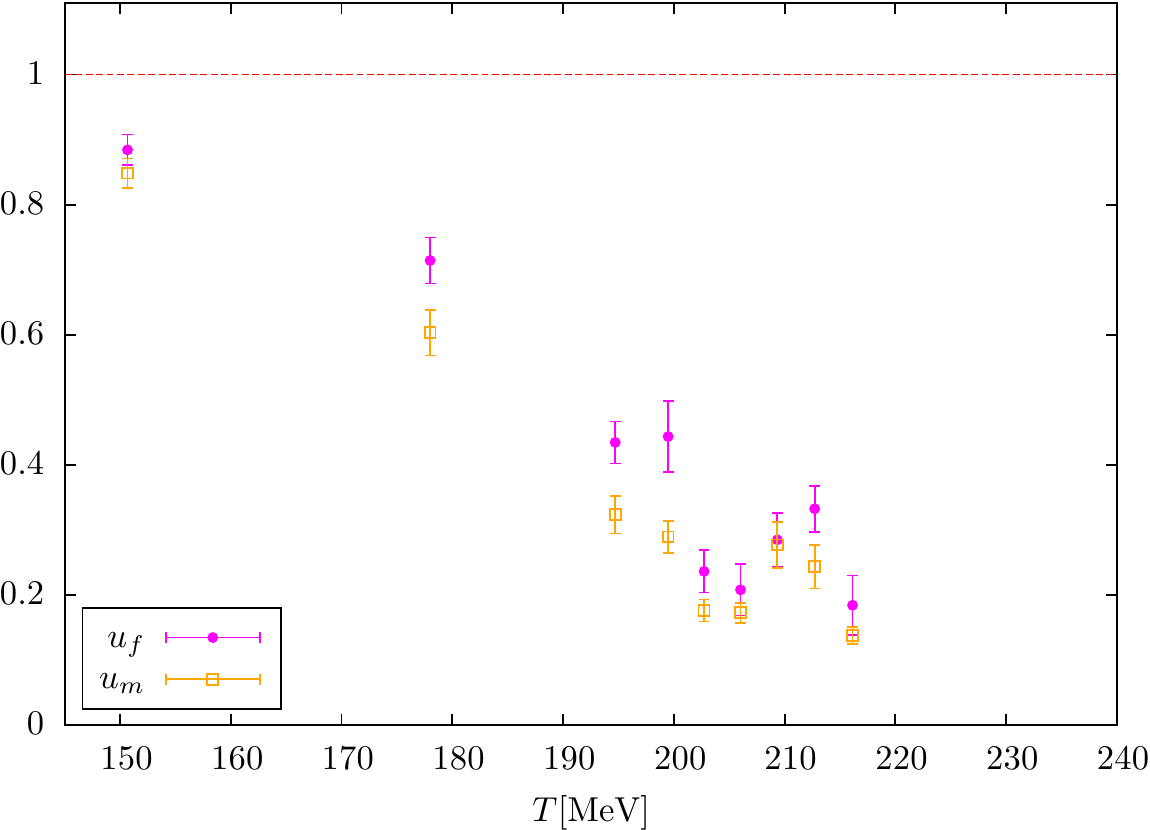}
    \includegraphics[width=.48\textwidth]{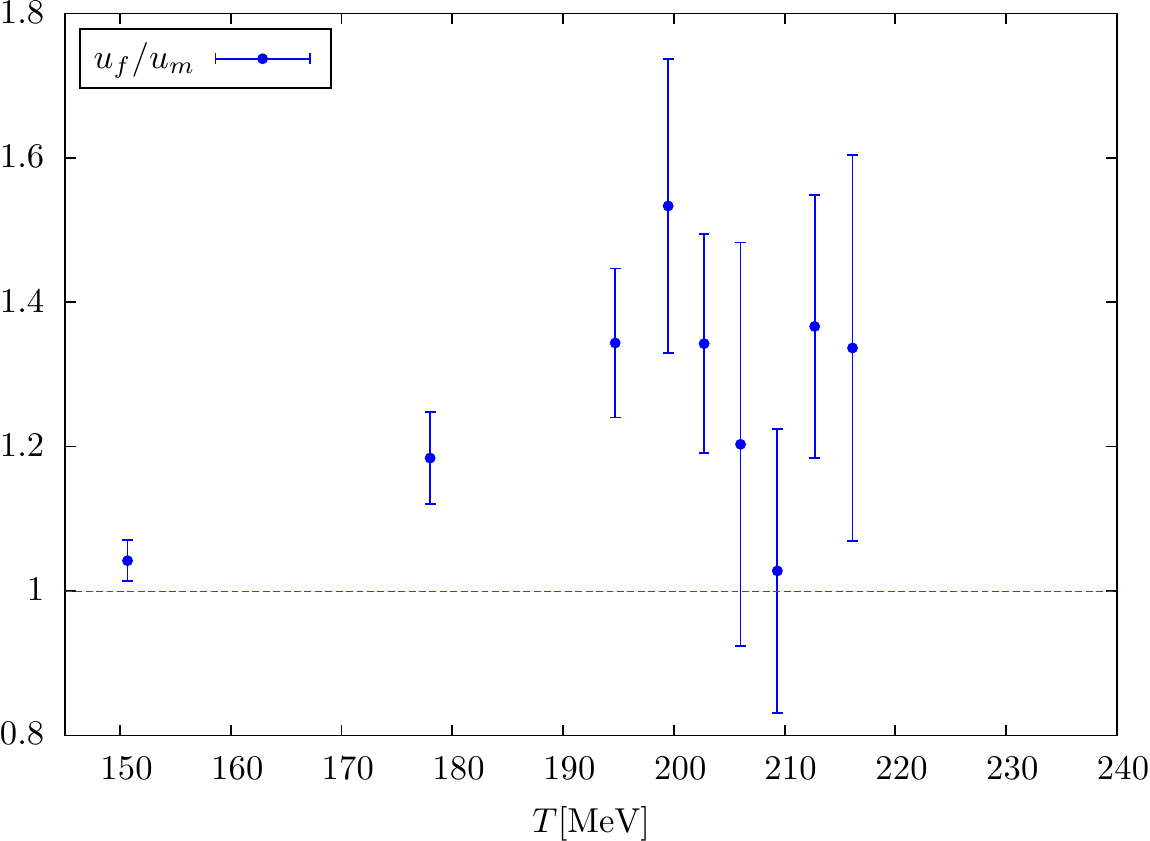}
\end{center}
\caption{Left: The two estimators of the pion velocity in the C1 scan. 
Right: Ratio of the estimators, which serves as a test of the chiral prediction 
(\ref{eq:Goopp}).}
\la{fig:u}
\end{figure}

We showed that, at sufficiently small quark mass,
the axial charge correlator is dominated by 
a light quasiparticle and that its mass $\omega_{\vec 0}$ 
is given by \eq(\ref{eq:w0simple}).
With $\omega_{\bf 0} = u m_\pi$, the following estimator for $u$ can be defined,
\be
u^2_m = -\frac{4m^2}{m^2_\pi}\left.\frac{G_{\rm P}(x_0, {\bf 0})}{G_{\rm A}(x_0, {\bf 0})}\right|_{x_0 = \beta/2}.
\ee
We introduce a second estimator for $u$ based on \eq(\ref{eq:ush}), 
\be
u_f \sinh(u_f m_\pi \beta/2) = \frac{f^2_\pi m_\pi}{2G_{\rm A}(\beta/2, {\bf 0})}.
\ee
It should be noticed that the pion velocity is a renormalization group invariant
quantity and thus, does not require any renormalization. The results for $u_f$ and $u_m$
are shown in \fig\ref{fig:u}. We observe a significant reduction of both quantities from unity,
pointing to a pion `velocity' well below the speed of light. However, whether the interpretation 
is valid for $T\gtrsim160\,{\rm MeV}$ is questionable. 

One way to test the validity of the chiral effective theory
predictions is the following.  The chiral EFT makes a prediction for
$G_{\rm P}(\beta/2)$ in terms of $f_\pi$, $m_\pi$ and $G_{\rm
  A}(\beta/2)$; see \eq(\ref{eq:w0simple}--\ref{eq:Goopp}).  Testing
whether $u_f/u_m=1$ is equivalent to testing this prediction.  It is
worth noting that at high temperatures, well in the deconfined phase,
$u_m = {\rm O}(m^2/T^2)$, while $u_f = {\rm O}(m/T)$, so that
$u_f/u_m$ is expected to grow with temperature. In the lattice data
displayed in the right panel of \fig\ref{fig:u} we indeed observe that
$u_f/u_m$ grows above unity. Thus it is at the lowest-temperature
ensemble in the C1 temperature scan that we are most confident in the
interpretation of $u_f$ as the pion quasiparticle velocity.

\subsection{The $T=0$ ensemble and test of chiral perturbation theory predictions}

In addition to the analysis of thermal ensembles, it is interesting to
compute the same observables on a corresponding zero-temperature
ensemble.  One reason is that we obtain the reference values of
$\omega_{\vec 0}$, $m_\pi$ and $f_\pi$ at $T=0$; the thermal
modification of these quantities can be compared with the predictions
of chiral perturbation theory~\cite{Gasser:1986vb,Gerber:1988tt}. A
second, practical reason is to check the validity of our estimators for
$u(T)$, since $\lim_{T\to0} u(T)=1$. We therefore analyze the CLS
ensemble labelled A5 in \cite{Fritzsch:2012wq}. All ensemble
parameters coincide with the lowest-temperature ensemble in the C1
scan; the only difference is the lattice extent in the time direction,
which is 64 instead of 16. The bare parameters and the computed
observables are summarized in \tab\ref{tab:A5pars}.

In contrast to the thermal ensembles, here we are able to directly
extract the mass of the pion propagating in the temporal direction,
which we denote by $\omega_{\bf 0}$. It is extracted by fitting to a
constant the coshmass of the pseudoscalar-pseudoscalar correlator,
where a clear plateau is observed. The pion decay constant $f_{\pi,
  \bf{0}}$ is calculated by fitting the amplitude of the axial charge correlator 
$G_{\rm A}(x_0)$. The effective quark condensate $|\left<\bar
{\psi} \psi \right>^{_{\overline {\rm MS}}}_{\rm GOR, \bf
  0}|^{1/3}$ given in \tab\ref{tab:A5pars} follows the definition (\ref{eq:CondGOR}), 
except that $m_\pi$ was replaced by $\omega_{\bf 0}$ and $f_{\pi}$ by $f_{\pi,  \bf{0}}$.

We find that the extraction of the pseudoscalar mass in the spatial
and in the temporal direction give the same answer within two standard
deviations. The estimators $u_f$ and $u_m$ are both compatible with
unity within two standard deviations; this adds to our confidence that
the estimators work as expected in practice.

We use $m_\pi$ and $f_\pi$ to normalize the corresponding quantities
at finite temperature in \fig\ref{fig:mfc}. This allows for the most
natural comparison of the predictions of one-loop chiral perturbation
theory~\cite{Gasser:1986vb}, an expansion around $(T=0,m=0)$, with the
lattice data. We display both the prediction for the infinite-volume
system and for the finite-volume system; details are given in appendix
\ref{sec:apdxGL}. At the lowest temperature in the C1 scan ($T\simeq
150\,{\rm MeV}$), the prediction agrees very well with the lattice
result for $f_\pi$. The central value of the correlation length
$m_\pi^{-1}$ lies somewhat lower than the corresponding chiral
prediction, but still within two standard deviations. However, on the
next ensemble, at $T\simeq 177\,{\rm MeV}$, the lattice data clearly
deviates from the chiral prediction. From this temperature onwards, both
$m_\pi$ and $f_\pi$ deviate substantially from their $T=0$
counterparts.  One-loop chiral perturbation theory predictions at $T\gtrsim
170\,{\rm MeV}$ appear to be unreliable.

We remark that the prediction for $f_\pi(T)/f_\pi(0)$ does not involve
directly the relation between the quark mass and the pion screening
mass.  The GOR-condensate, however, does; it is compared to the chiral
prediction in \fig\ref{fig:c2}. Here the quantities combine to give a
result which is only mildly temperature-dependent. Correspondingly the
chiral prediction lies numerically quite close to the data points. The
prediction for the GOR condensate seems to be more robust than the
predictions for $m_\pi$ and $f_\pi$ taken separately; it works, at our
current level of accuracy, essentially up to the transition
temperature.

We can in principle compare the pion quasiparticle mass, 
computed as $\omega_{\vec 0}(T) = u(T) m_\pi(T)$, with the two-loop predictions
of chiral perturbation theory~\cite{Schenk:1993ru,Toublan:1997rr}. 
At $T=150\,{\rm MeV}$ in the C1 scan, we find
\be
\frac{\omega_{\vec 0}(T)}{\omega_{\vec 0}(0)} = 0.97(4), 
\ee
where $\omega_{\vec 0}(T=0)= 294(4)\,{\rm MeV}$. Thus the thermal shift of the pion quasiparticle
mass appears to be very small.
At the same temperature, but at the physical quark mass, the corresponding quantity is predicted 
to be about 0.86 at the two-loop level~\cite{Schenk:1993ru}; we note that there is a change 
in the sign of $\frac{\omega_{\vec 0}(T)}{\omega_{\vec 0}(0)} -1$ 
between the one-loop and the two-loop result at this temperature.
Clearly one expects the thermal effect on $\omega_{\vec 0}$ 
to be smaller at heavier quark masses. Thus the lattice results are not obviously inconsistent 
with the chiral prediction. We postpone a more detailed comparison of lattice results 
for the quantity $\omega_{\vec 0}$ (and indeed $\omega_{\vec k}$)  
with chiral perturbation theory to a future study.

The relative success of the one-loop chiral prediction for the thermal effect on 
the `chiral' quantities $m_\pi$, $f_\pi$ and $\<\bar\psi\psi\>_{\rm GOR}$ at 
\be
T=150\,{\rm MeV} \simeq 0.7 T_c \qquad (\overline{m}^{\overline{\rm MS}}\simeq 15\,{\rm MeV}) 
\ee
is somewhat unexpected when one considers that the energy density,
say, for physical quark masses is completely dominated by hadrons more
massive than pions~\cite{Gerber:1988tt,Borsanyi:2013bia}.  The
surprise at the quark mass used here is, in a sense, that chiral
quantities are still affected below the $10\%$ level by the thermal
effects. However, the effect of the thermal medium increases rapidly
above $T=150\,{\rm MeV}$.

\subsection{Quark mass dependence of $m_\pi$ and $f_\pi$ around the pseudocritical temperature}

The scan D1 at the light quark mass is more concentrated around the
pseudocritical temperature. Therefore we can only discuss the quark
mass dependence of the observables discussed so far in the crossover
region; see \fig\ref{fig:mpifpi_mq}. We find that, if $m_\pi/T$ and
$f_\pi/T$ are viewed as a function of $T/T_c$, where $T_c$ is the
quark-mass dependent pseudocritical temperature, the quark mass
dependence is very mild. In this respect we are far from the deeply
chiral regime where the screening pion mass exhibits a sudden steep
rise at $T_c$, from a low value below $T_c$ of order $\sqrt{m}$.

A look at \tab\ref{tab:C1summ}--\ref{tab:D1summ} shows that 
the effective condensate appears to be quite insensitive to the
quark mass up to $T=195\,{\rm MeV}$, which corresponds to the
pseudocritical temperature at the lower quark mass. In other words,
the GOR relation is satisfied within the uncertainties, in spite of
the fact that, at fixed temperature, the pion mass $m_\pi$ does not decrease with the quark
mass  between scan C1 and scan D1.

\begin{figure}[t!]
\includegraphics[width=.48\textwidth]{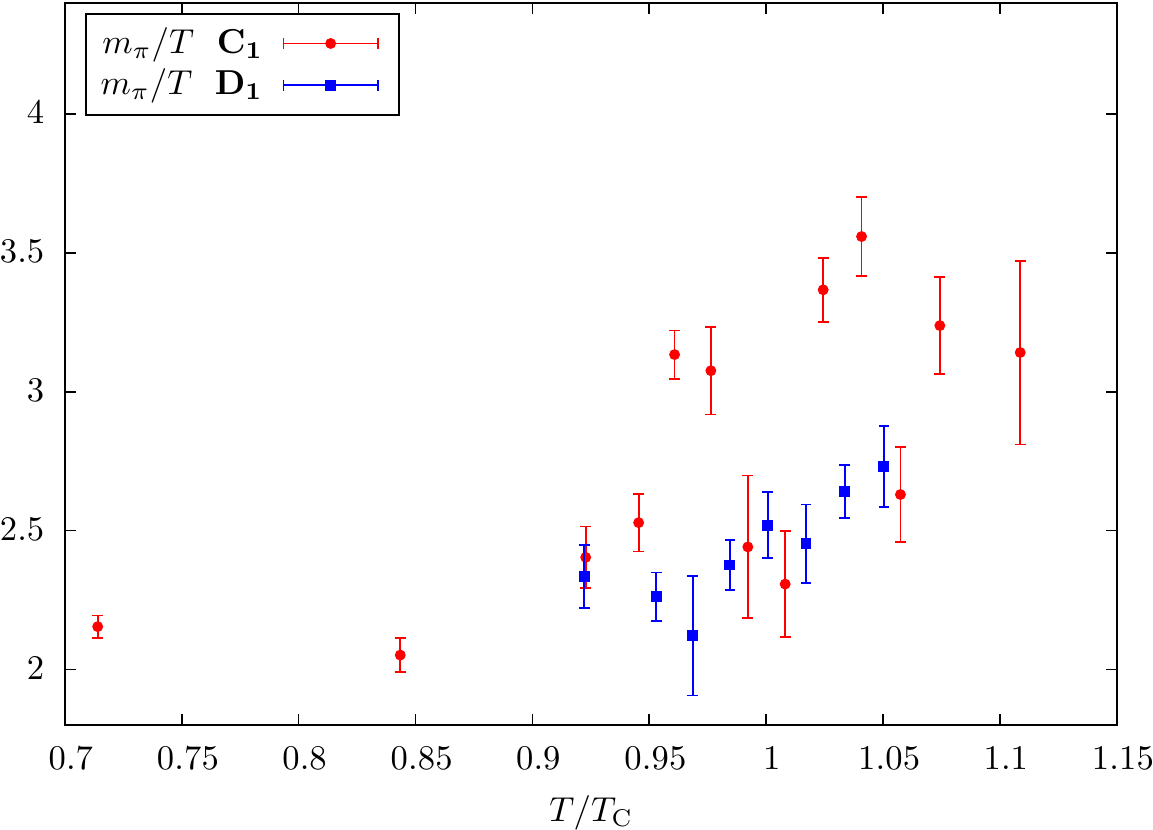}
\includegraphics[width=.48\textwidth]{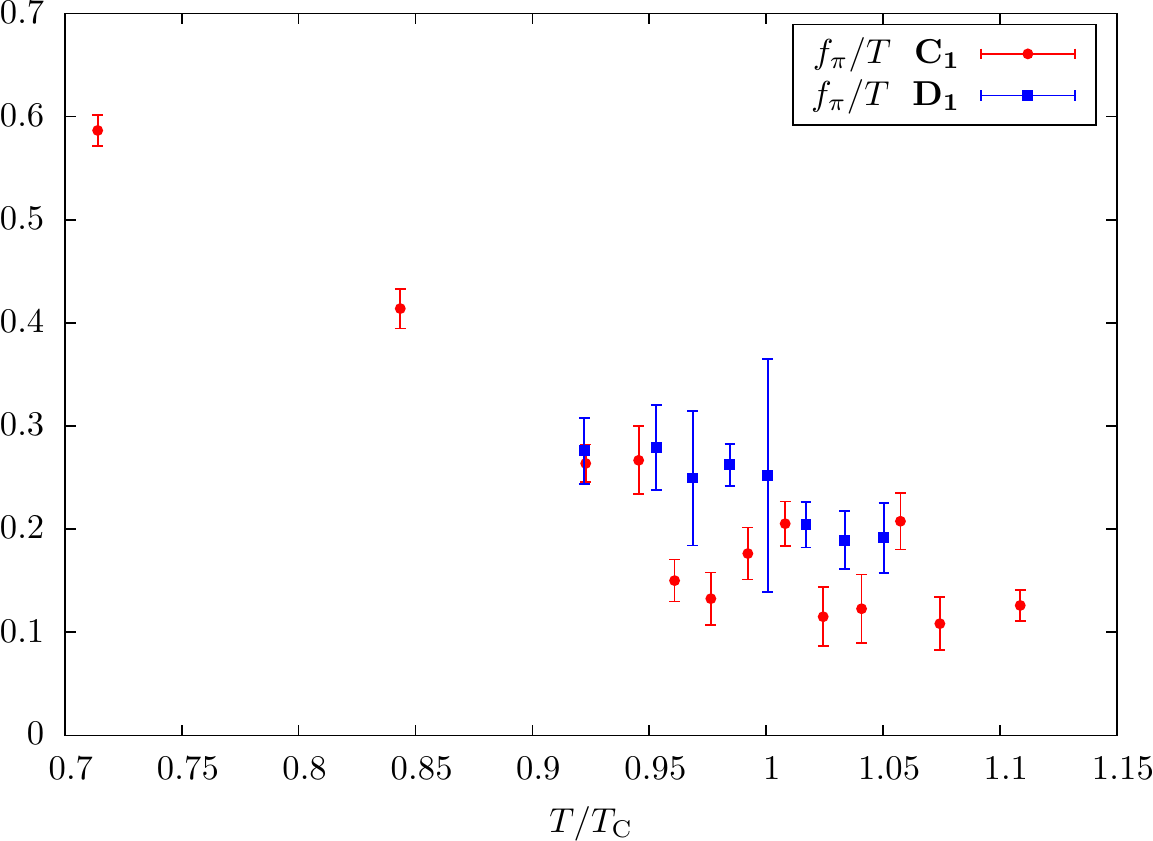}
\caption{Temperature dependence of the pion screening mass and the associated decay constant, for two quark masses. 
The temperature is given in units of the pseudocritical temperature at the corresponding quark mass.}
\la{fig:mpifpi_mq}
\end{figure}

\section{Spectral function reconstruction using the maximum entropy method\la{sec:mem}}

So far we have concentrated on computing the properties of the pion
quasiparticle indirectly from spatial correlation functions, relying
on the chiral effective theory. Given the lattice extent is larger in
the spatial directions, this approach has the advantage that masses
and amplitudes can be calculated quite accurately from the data. For
example determining the pion pole mass from the temporal correlation
function is not possible given only $N_t/2=8$ points on the available
lattice ensembles, while we achieve an accuracy of roughly $4\%$ on
the same ensembles in the spatial direction, i.e. with $N_s/2=16$
points. Since all quantities of interest here are accessible from the
spectral function, an alternative approach is to study the behavior of
the spectral functions underlying the temporal correlation functions.

To achieve this one has to invert the kernel $K(x_0,\omega)=\frac{\cosh(\omega(\beta/2-x_0))}{\sinh(\omega\beta/2)}$, 
see Eqs.~(\ref{eq:SF1})-(\ref{eq:SFA}).
Inverting this type of equation in order to extract the spectral
function is a typical ill-posed problem. One commonly adopted
procedure to compute spectral functions from lattice correlation
functions is the Maximum Entropy Method (MEM) 
\cite{Asakawa:2000tr,Nakahara:1999vy,Yamazaki:2001er,Fiebig:2002sp,Sasaki:2005ap,Karsch:2001uw,Aarts:2007wj,Aarts:2007pk}. 
In this method the guiding principle for the selection of the most likely
solution given the finite number of lattice data points with errors and
an input default model is Bayesian statistical inference.

We adopt the implementation of MEM presented
in \cite{Ding:2010ga,Ding:2011hr,Ding:2012sp}, which is based on
Brian's algorithm. 
Defining the modified kernel 
\ba\la{eq:K}
\tilde{K}(x_0,\omega)&\equiv &\tanh(\omega\beta/2) K(x_0,\omega),
\ea
the spectral function is parametrized as 
\be
\rho(\omega) = m(\omega)\exp(f(\omega)),
\ee
where $m(\omega)$ is an input default model and
 $f(\omega)$ is expanded in a basis of functions that depends on the choice
of the kernel $\tilde K$. 
Due to the divergence of $K(x_0,\omega)$ as $\omega\rightarrow 0$, the
redefinition (\ref{eq:K}) ensures a stable behavior of MEM around $\omega\sim 0$,
while retaining the large frequency behavior of the original
kernel. This is one choice for the modified kernel, however different
redefinitions are possible and have been used in the
past \cite{Aarts:2007wj,Aarts:2007pk}.

The choice of input default model plays a
crucial role in any spectral function reconstruction using MEM and
currently poses the largest source of error. It enters into the
definition of the Shannon-Jaynes entropy term
\be
S[\rho]=\alpha \int_0^\infty  \frac{d\omega}{2\pi}\, \Big[\rho(\omega)
 - m(\omega) - \rho(\omega) \log \Big( \frac{\rho(\omega)}{m(\omega)} \Big)\Big],
\ee
and, in the Bayesian language, is part of the prior information $H$. 
The parameter $\alpha$ weights the relative importance of the data and the prior knowledge.
Given this term and the standard likelihood function $L[\rho]=\chi^2/2$, 
the most probable spectral function $\rho(\omega)$ underlying the lattice correlator $G$ 
can be obtained by maximizing the conditional probability
\be
P[\rho|GH] = \exp( S[\rho] - L[\rho] ).
\ee
There is a lot of freedom in choosing the default model $m(\omega)$ for the
entropy term. However, it can be shown \cite{Asakawa:2000tr} that,
given precise enough data, MEM will produce a unique solution, if it
exists, regardless of the default model. Unfortunately, in practice
the data is not accurate enough to ensure this property and it is not
a priori clear how a specific choice of default model impacts the
obtained solution. Therefore great care must be taken in MEM analyses to
check the dependence on the default model by repeating the analysis
with several sufficiently different classes of them. Any stable features accross all
results should then be safe to interpret in terms of physics.

\subsection{Choice of default models and vacuum spectral functions}

In this study we will choose default models corresponding to
the infinite temperature limit on the one hand and the zero temperature
case on the other.  The former corresponds to a system of
non-interacting quarks, and the analytically known spectral
functions \cite{Karsch:2003wy,Aarts:2005hg} provide a default model. 
Specifically we choose
\be
m^{\rm free}_{_{\rm P}}(\omega)\sim \omega^2\tanh(\omega/2T) \quad \textrm{ and }\quad
m^{\rm free}_{_{\rm A}}(\omega)\sim \tanh(\omega/2T),
\la{eq:freedm}
\ee
even though the proportionality constants are known analytically, in
the actual analysis we freely vary them as an additional
crosscheck. These default models are essentially featureless and leave
the most `freedom' to the MEM analysis to extract excitations. For the zero temperature
limit the A5 lattice ensemble plays a crucial role, as it enables the
reconstruction of vacuum spectral functions on a large lattice with
very accurate data.  We therefore reconstruct the spectral functions
$\rho^{\rm vac}_{_{\rm P}}(\omega)$ and $\rho^{\rm vac}_{_{\rm A}}(\omega)$ using the free
default models Eq.~(\ref{eq:freedm}) and subsequently define them to
be the `vacuum' default models
\be
m^{\rm vac}_{_{\rm P}}(\omega)=\rho^{\rm vac}_{_{\rm P}}(\omega) \quad \textrm{ and }\quad
m^{\rm vac}_{_{\rm A}}(\omega)=\rho^{\rm vac}_{_{\rm A}}(\omega),
\la{eq:vacdm}
\ee
to be used in the analysis of the thermal correlators.

\begin{figure}[t]
\includegraphics[width=.49\textwidth]{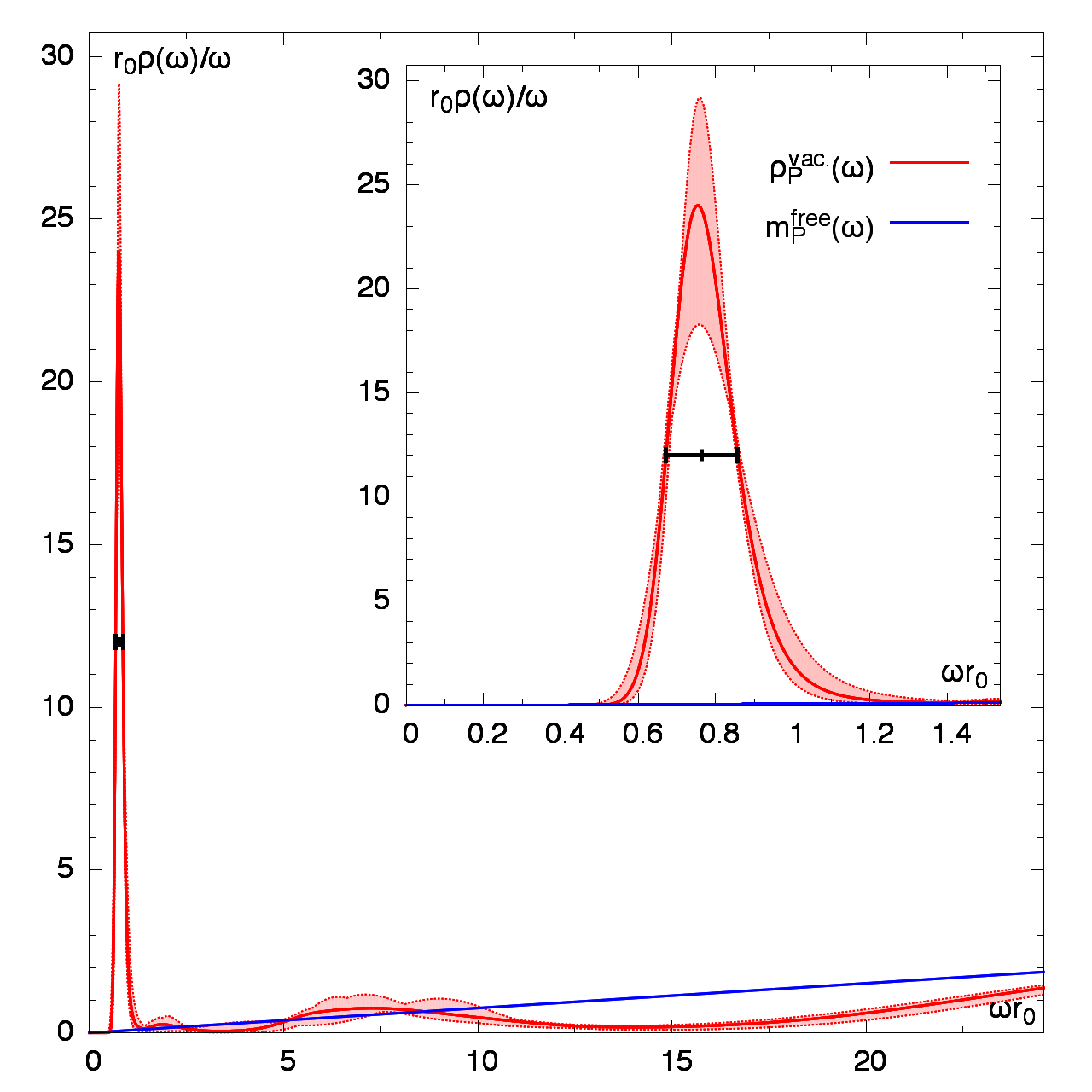}
\includegraphics[width=.49\textwidth]{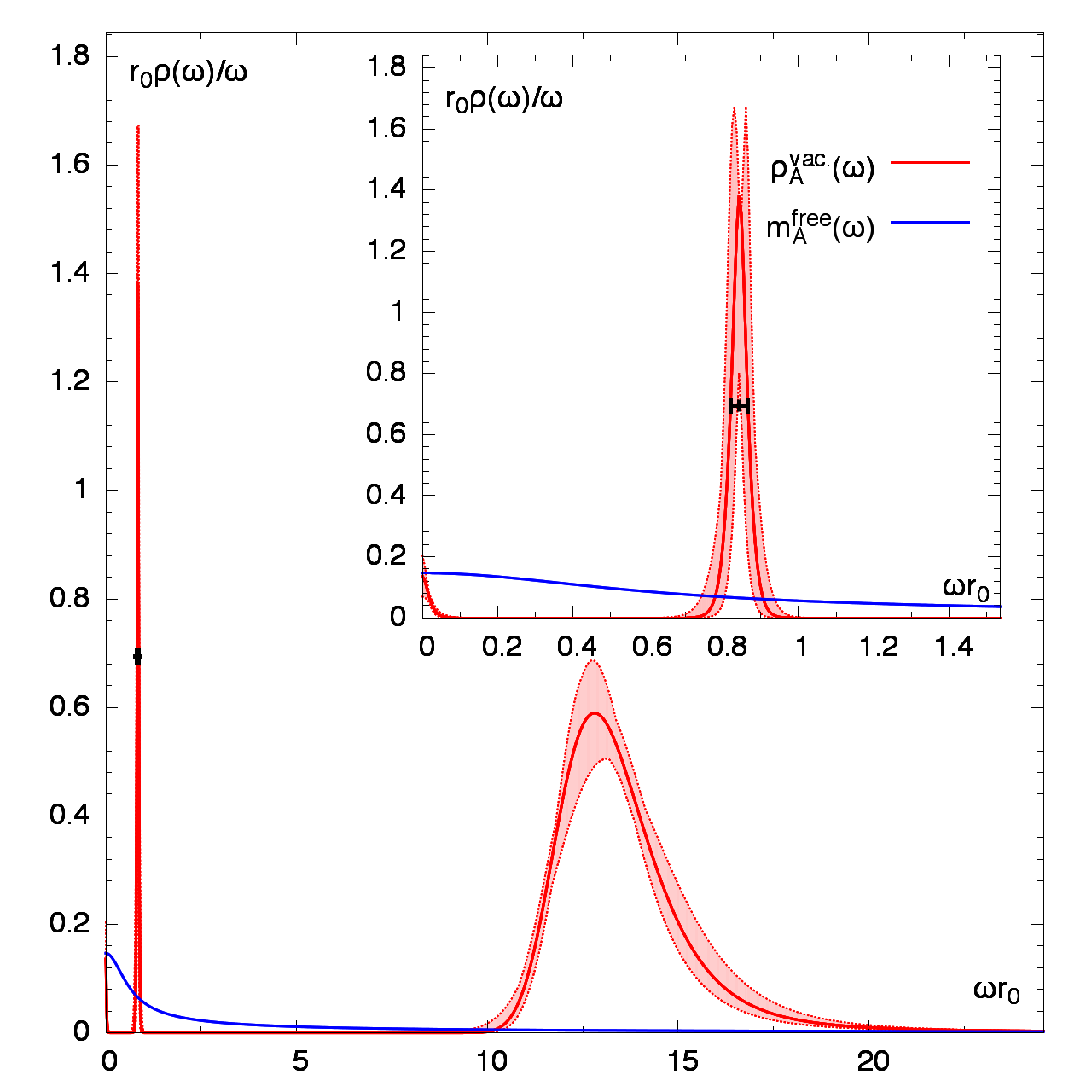}
\caption{{The MEM reconstruction of the vacuum $P$ (left) and $A_0$ (right)
channel spectral functions. In both cases a free theory inspired
default model $m_{_{\rm A/P}}^{\rm free}(\omega)$ is used as prior information
for the MEM analysis. The black error bars indicate the half-maximum
width and midpoint, which we define to give $m_\pi$ and its error. The
broad peak structure at large frequencies is understood as lattice
artifact. The quoted error bands represent the spread of spectral
functions obtained in a jackknife analysis.}}
\label{fig:spf-vacuum}
\end{figure}

In Fig.~\ref{fig:spf-vacuum} we show the resulting spectral functions
over frequency in units of the reference scale $r_0$, along
with their respective free default models in the $P$ (left) and the
$A_0$ (right) channels. To estimate the statistical uncertainties, the
MEM analysis is repeated on a set of jackknife samples, the given
error band shows the spread of the resulting spectral functions.  In
both cases we observe the emergence of a narrow peak in the low
frequency region and a broad peak structure at large frequencies. The
second peak structure is to be understood as a lattice artifact, as
the free lattice spectral functions also exhibit such a
structure \cite{Aarts:2005hg}.  In both channels we observe a clear
separation between the low frequency spectrum dominated region and the
lattice cut-off region. In the $P$ case this separation is located
roughly around $\omega a\sim 0.5-1.0$, while in the $A_0$ channel we
observe a separation window from $\omega a\sim 0.25$ through to
$\omega a\sim 1.5$.  In the next step we associate the low frequency
peak structure with the pion.  To read off its mass from the spectral
functions obtained by the MEM analysis, we calculate the peak maximum
on the unsampled result. The mass and `resolution error' are then
given by the width of the peak at half-maximum and its midpoint (black
error bars in Fig.~\ref{fig:spf-vacuum}).  Combining the resulting
values with the lattice spacing $a=0.0818$fm we obtain
$m_{\pi}^{P}=300(36)\,{\rm MeV}$ and $m_{\pi}^{A}=331(9)\,{\rm MeV}$.  Camparing these
values to those obtained from fitting the spatial and temporal
correlators in the $P$ channel tabulated in \tab\ref{tab:A5pars},
we find very good agreement for the result originating from the $P$
channel, while the result extracted from the $A_0$ channel is
larger. The likely explanation is the observed peak located around
$\omega\simeq 0$ in the $A_0$ channel in Fig.~\ref{fig:spf-vacuum}
(right).  This contribution to the spectral function is then
compensated, in the MEM reconstruction, by a shift of the pion peak
position to larger frequencies.

The clear statistical stability of the vacuum spectral functions
obtained on the A5 lattice ensemble using the free default models in
Fig.~\ref{fig:spf-vacuum}, motivates us to choose the average
spectral function (red lines in Fig.~\ref{fig:spf-vacuum}) as default
models for the MEM reconstruction of the finite temperature spectral
functions.

\subsection{Thermal spectral functions in the $A_0$ channel and $f^2_\pi/u^2$}

\begin{figure}[t]
\includegraphics[width=1.0\textwidth]{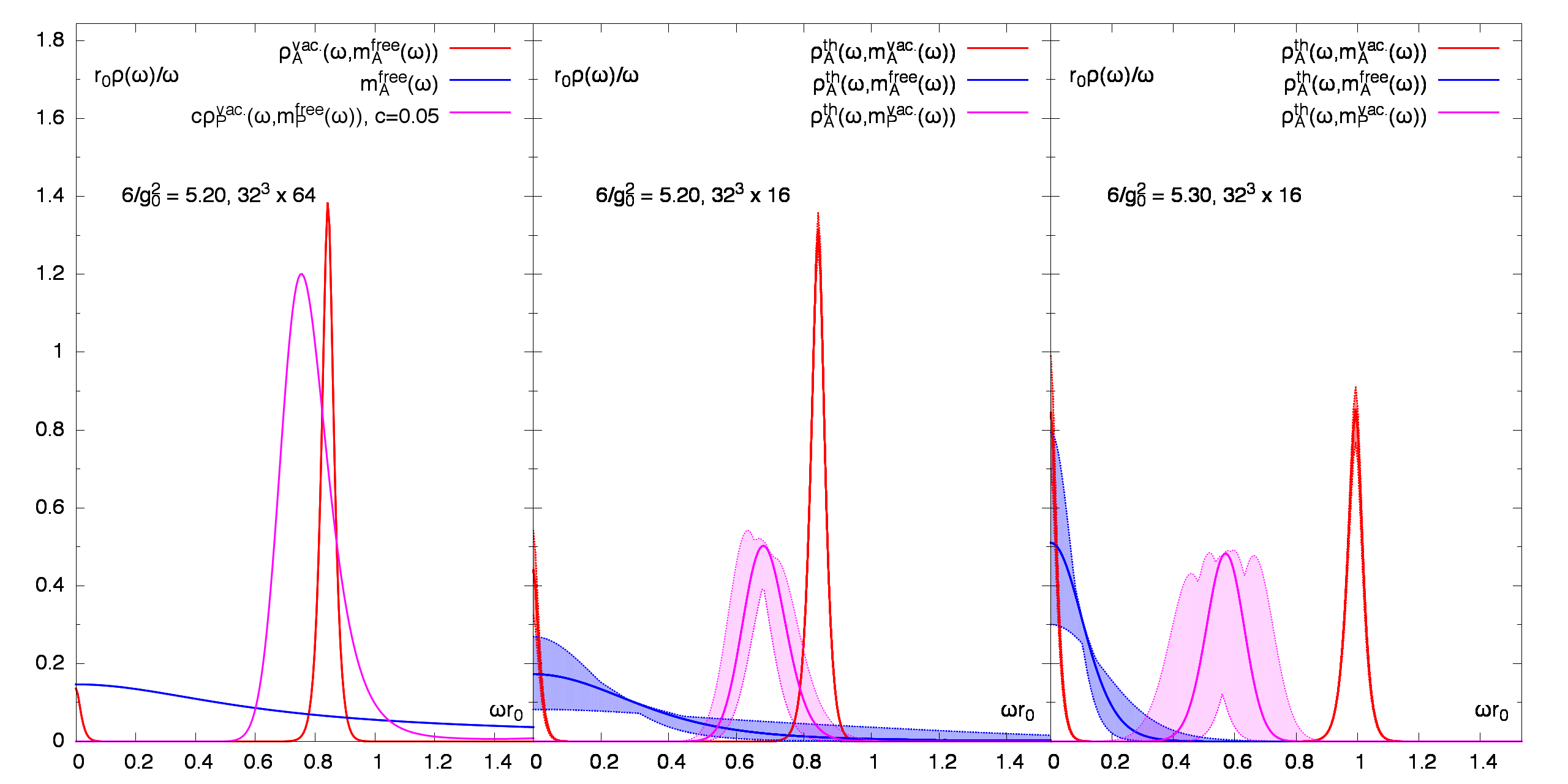}
\caption{{Reconstruction of the spectral function $\rho_{_{\rm A}}$ of the axial charge
at finite temperature and vanishing spatial momentum. The notation $\rho_{_{\rm A}}= \rho^A_T(\omega,m(\omega))$
emphasizes the default-model dependence.
Left: 
The three input default models used for the reconstruction of the thermal 
spectral function. 
Middle:
The MEM reconstruction of the spectral function 
on the $6/g_0^2=5.20$  ensemble of the C1 scan
for all three default models. The quoted error bands represent the spread 
of spectral functions obtained in a jackknife analysis. 
Right: 
The corresponding MEM results for the $6/g_0^2=5.30$ ensemble 
of the C1 scan.}}
\label{fig:spf-fixedscale}
\end{figure}

Following the discussion of Sec.~\ref{sec:spf}, using
Eqs.~(\ref{eq:CP}) and (\ref{eq:rhoA}) the integral over the spectral
function in the $A_0$ channel can be linked to $f_\pi^2/u^2$,
\be
\rho_{_{\rm A}}(\omega)=\frac{f_\pi^2 m_\pi}{2u}\delta(\omega-\omega_0)\quad\Longrightarrow\quad
{\cal A}(\Lambda)\equiv 2\int_0^\Lambda \frac{d\omega}{\omega} \rho_{_{\rm A}}(\omega) = \frac{f_\pi^2}{u^2},
\la{eq:area}
\ee
where $\Lambda$ is a scale separation parameter between the low and
large frequency regions.  While the details of the spectral functions
themselves, like peak positions and widths, are generally very
sensitive to the input default model, the area under these spectral
functions in a given interval is more robust.

In the following, we reconstruct the thermal spectral functions in the
$A_0$ channel and compute the area according to
\eq(\ref{eq:area}). For every available ensemble we choose
$m^{\rm free}_{_{\rm A}}(\omega)$, $m^{\rm vac}_{_{\rm A}}(\omega)$ and
$m^{\rm vac}_{_{\rm P}}(\omega)$ as input default models. Using these three
models we cover a range of inputs that go from a very smooth, featureless
($m^{\rm free}_{_{\rm A}}(\omega)$) to a rather specific
($m^{\rm vac}_{_{\rm A}}(\omega)$) model. We show the default models in the
left panel of Fig.~\ref{fig:spf-fixedscale}, whereby we rescaled
$m^{\rm vac}_{_{\rm P}}(\omega)=\rho^{\rm vac}_{_{\rm P}}(\omega,m^{\rm free}_{_{\rm P}}(\omega))$ by a
factor $1/c=20$ for readability.  In the middle and
the right panel, we show the resulting thermal spectral functions
$r_0\rho_{_{\rm A}}(\omega,\vec 0)/\omega$ and their statistical error bands for the
$6/g_0^2=5.20$ and $6/g_0^2=5.30$ ensembles. Note once more that the bare parameters 
of the $6/g_0^2=5.20$ ensemble of the C1 scan are identical to those of the $N_t=64$
reference ensemble. Already in this case we observe the free default model
does not lead to a sharp peak result in the finite temperature
ensemble. Instead the resulting spectral function exhibits a broad
peak centered around $\omega\simeq 0$. Reconstructing the spectral
function using the peaked default models on the other hand leads to
low frequency peaks, as observed in the vacuum case.  Comparing the
results at $6/g_0^2=5.20$ and $6/g_0^2=5.30$ we observe a drop in the peak
amplitudes for the peak-type default models; and a more narrow peak
for the results obtained from the free default model.

\begin{figure}[t]
\includegraphics[width=0.49\textwidth]{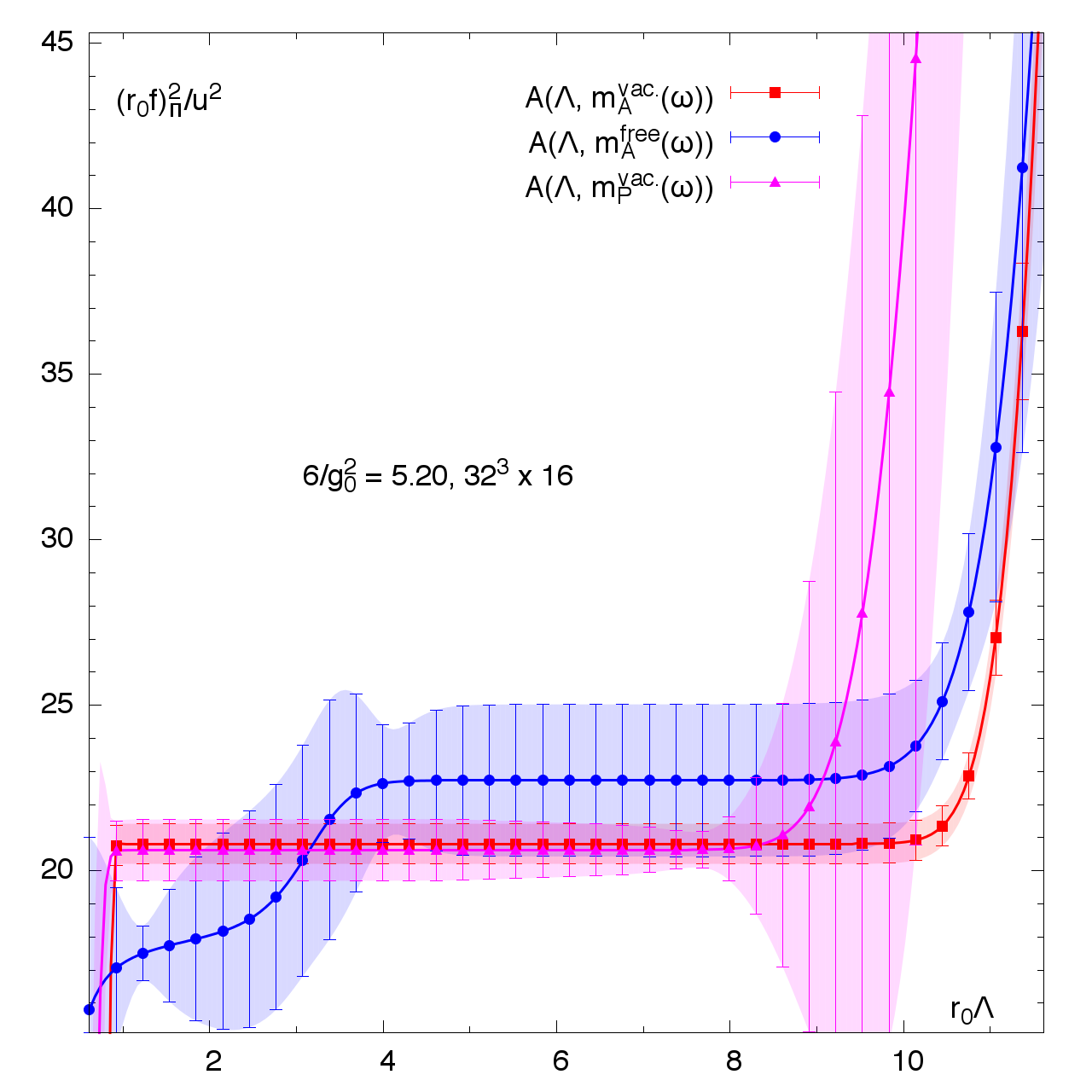}
\includegraphics[width=0.49\textwidth]{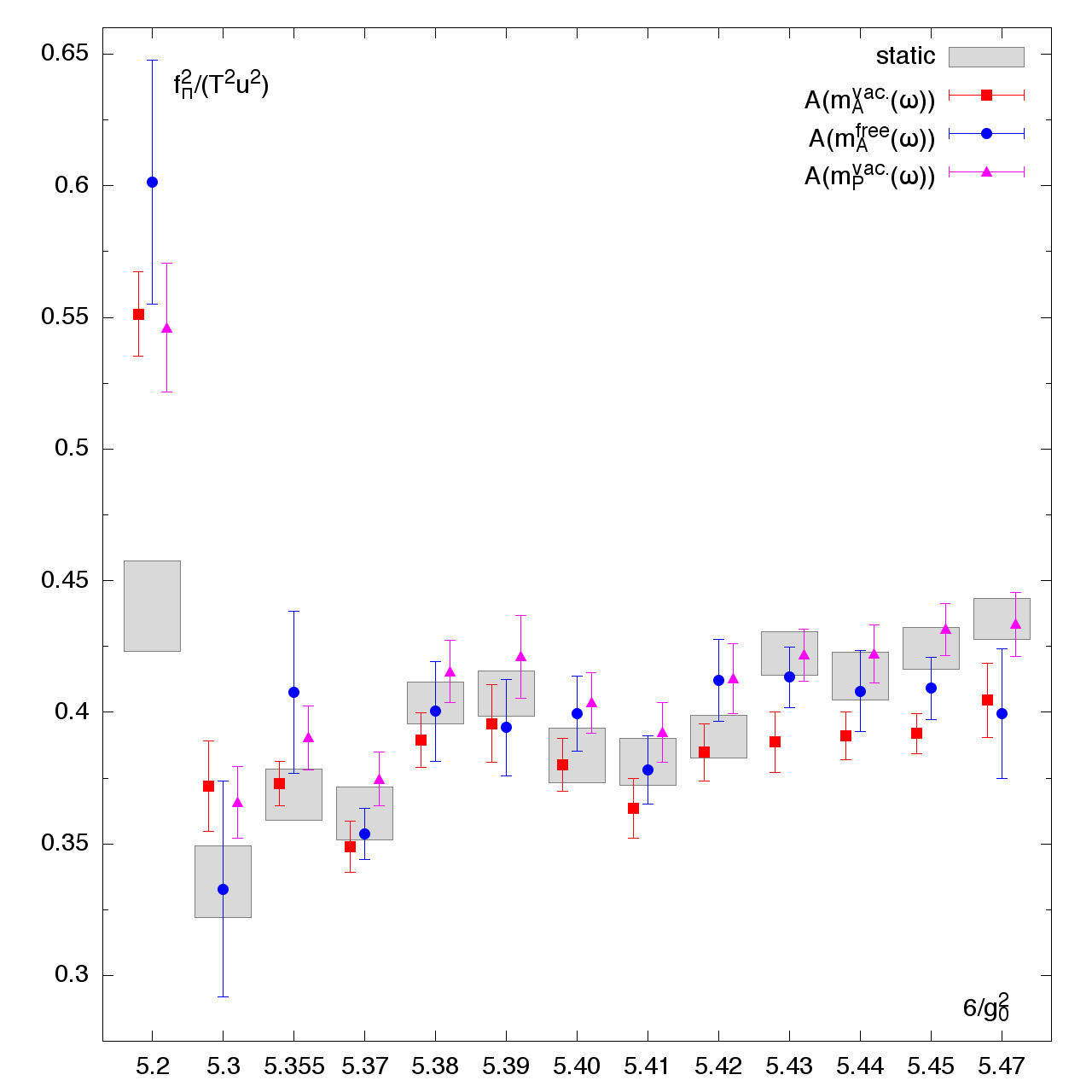}
\caption{{Left: The area ${\cal A}(\Lambda,m(\omega))$ for different values
 of the cut-parameter $\Lambda$ on the $6/g_0^2=5.20$ ensemble
of the C1 scan obtained from a MEM reconstruction using all three default models $m(\omega)$. 
We observe a clear plateau and therefore separation region. Right: 
The area ${\cal A}(\Lambda,m(\omega))$ for all available lattice ensembles in units of $T^2$, 
compared to the results of Sec.~\ref{sec:lat} for $f_\pi^2/(T^2u_f^2)$, which make use of
 static quantities.}}
\label{fig:area-cut}
\end{figure}

In the next step we compute the area under the spectral functions as
prescribed in \eq(\ref{eq:area}). To do so we first analyze the
dependence on the cut-off parameter $\Lambda$ at $6/g_0^2=5.20$ and
$N_t=16$, the result is shown in Fig.~\ref{fig:area-cut}(left). The
errors shown in this plot originate from a jackknife procedure to
calculate the integral over the spectral function up to the cut
parameter $\Lambda$.  For the two peak-type default models we observe
a clear plateau, i.e. cutoff-independence, for values of $r_0\Lambda$ above
about $1.2$, which in both cases is slightly above the peak
reagion. This plateau is seen to be stable up until
$r_0\Lambda\simeq 8$ for the $P$-type and $r_0\Lambda\simeq10$ for the
$A_0$-type default model.  For the free default model we observe a
cut-dependence up to values of $r_0\Lambda\simeq4.3$, the subsequent
plateau is stable up to $r_0\Lambda\simeq10$. In the following, we
choose a suitable value for $\Lambda$ by determining a local minimum
of the spectral function in the interval $ 1.2 \lesssim r_0\Lambda \lesssim10$.

Using the same default models (in units of temperature), 
we repeat the MEM-reconstruction and subsequent determination of
$f_\pi^2/u^2$ for all available lattice ensembles. Rescaling the
latter quantity by $T^2$, we compare the results in the C1 scan with those
obtained using static quantities in Fig.~\ref{fig:area-cut} (right
panel). As before, the quoted errors originate from a jackknife
analysis in the MEM reconstruction and we use the three default models
described above. As an example, we give our MEM results for
${f_\pi/u}$ on the $N_t=64$ and $N_t=16$ ensembles at $6/g_0^2=5.20$,
\be
\left(\frac{{f}_\pi}{u}\right)^{T=37{\rm MeV}}=110(23){\rm MeV} \quad {\rm and}\quad
 \left(\frac{{f}_\pi}{u}\right)^{T=150{\rm MeV}}=113(19)(32)(24)\rm{MeV},
\ee
where we quote the average value over all MEM-results and the errors
from the $A_0$-vacuum, $A_0$-free and $P$-vacuum default models
respectively. 

The grey shaded areas in Fig.~\ref{fig:area-cut} (right) denote the
results obtained from the analysis of Sec.~\ref{sec:lat}, which uses
static correlation functions. Overall we find good agreement between
the MEM based results and the approach of Sec.~\ref{sec:lat}.  In the
lowest-temperature ensemble of the C1 scan, the MEM results overshoot
the static results. A possible reason is that in Sec.~\ref{sec:lat}, 
the spectral function $\rho_{_{\rm A}}$ is assumed to be given by a 
single `delta function', while the result of the MEM reconstruction 
exhibits (for each default model) a more complicated spectral weight
distribution. It is at the lowest temperature that the comparison is most 
sensitive to this difference.
We note however that the agreement improves if one uses
the estimator $u_m$ instead of $u_f$ in the quantity $f_\pi/u$ computed in
Sec.~\ref{sec:lat}.

\section{Outlook\la{sec:outl}}

This work represents a step towards understanding the degrees of freedom
dictating the static correlations and the dynamical properties of QCD
in its low-temperature phase. We have computed the two temperature-dependent
parameters that determine the pion quasiparticle dispersion relation
(see \eq(\ref{eq:intro_disprel})). The results are compared to (mostly one-loop)
predictions of chiral perturbation theory.
The methods introduced in this paper
can be applied to ensembles with a quark content closer to the real
world: it would be very interesting to compute all observables
considered here on QCD ensembles with up, down and strange quarks at
their physical masses (see the recent \cite{Bhattacharya:2014ara}). In
order to test the functional form of the pion quasiparticle dispersion
relation (\ref{eq:intro_disprel}), it would be very interesting to analyze data at
non-vanishing spatial momentum $\vec k$. 

The results presented in this paper are still subject to cutoff and
finite volume effects, which have not been investigated. Especially
the latter can still be quite sizeable since the ensembles at our
disposal have a rather small spatial extent with an aspect ratio of
$LT=2$. Currently the set of ensembles is extended to aspect ratios of
$LT=3$ and 4. Cutoff effects are presumably small, since with $1/aT=16$ we
are at the state-of-the-art concerning the temporal extent at the
transition temperature. However, a systematic study of cutoff effects
would also be desirable.

In order to test the $T=0$ chiral effective theory predictions more
stringently, additional simulations at temperatures 100-150MeV are
required at lighter quark masses. The comparisons with the available
two-loops calculations in chiral perturbation
theory~\cite{Schenk:1993ru,Toublan:1997rr} could then be done
systematically.

\appendix 
\section{Chiral Ward identities for two-point functions \& a sum rule \la{sec:apdxWI}}

The isovector vector and axial-vector currents, as well as the pseudoscalar density were defined in \eq(\ref{eq:VAPdef}).
We use the Euclidean field theory method to derive the axial Ward identities~\cite{Luscher:1996jn}.
We assume that all chemical potentials are set to zero.
It is useful to recall some of the space-time transformation properties of these local operators.
Under the Euclidean time reversal tranformation $(x'_0 = - x_0,\;\vec x'=\vec x)$, we have
\be
A^a_0{}'(x') =  A^a_0(x), \qquad A^a_k{}'(x') = - A^a_k(x),
\qquad  P^a{}'(x') = -P^a(x),\qquad 
\ee
while under $(x'= -x)$
\be
  A^a_\mu{}'(x')= -A^a_\mu(x), \qquad P^a{}'(x')= P^a(x).
\ee
We also note that $V_\mu$ is odd under charge conjugation $C$, while $P$ and $A_\mu$ are even.

The variations of the quark and antiquark fields under an infinitesimal, isovector, axial phase rotation read
\ba
\delta_A^a\psi(x) &= {\txts\frac{1}{2}}\tau^a\gamma_5 \psi(x),\qquad \qquad 
\delta_A^a\bar\psi(x) &= \bar\psi(x)\gamma_5{\txts\frac{1}{2}}\tau^a.
\ea
They lead to the following transformation of the composite operators,
\be
\delta_A^a A_\mu^b(x) = -i\epsilon^{abc} V_\mu^c(x), \qquad 
\delta_A^a P^b(x) = \frac{\delta^{ab}}{2} \bar\psi\psi,
\ee
For an axial transformation parameter $\alpha^a(x)$, the variation of the action is given by~\cite{Luscher:1998pe}
\be
\delta S = \int d^4x \left( \partial_\mu \alpha(x)^a A_\mu^a(x) + \alpha(x)^a 2m\,P^a(x)\right).
\ee
In the path integral, the invariance of the integration measure under the transformation above leads to
$\< \delta{\cal O}\> = \< {\cal O} \delta S\>$. In particular, if ${\cal O}$ consists of one local field located
 at the point $y$,
\be
\alpha(y)\< \delta_A^a {\cal O}(y)\> = \< {\cal O}(y) \int d^4x \left( \partial_\mu \alpha(x) A_\mu^a(x) + \alpha(x) 2m\,P^a(x)\right)\>
\ee
In the following we set $\alpha(x)=e^{ikx}$ and consider several choices for ${\cal O}$.
Choosing ${\cal O} = A_\nu^b$, we obtain
\be\la{eq:WIAmom}
0 = ik_\mu \< A_\nu^b(0)\int d^4x\; e^{ikx} A_\mu^a(x)\>
+ 2m \int d^4x\; e^{ikx}\< A_\nu^b(0) P^a(x)\>.
\ee
Choosing instead ${\cal O} = P^b$, we obtain
\be\la{eq:WIPmom}
\frac{1}{2}\delta^{ab}\<\bar\psi\psi\> = ik_\mu\<P^b(0) \int d^4x\; e^{ikx} A_\mu^a(x)\> + 2m\<P^b(0)\int d^4x\;e^{ikx}P^a(x)\>.
\ee
These are the momentum-space versions of the Ward identities, 
while \eq(\ref{eq:WIA}--\ref{eq:WIP}) are the position-space versions.

\subsection{A sum rule for the spectral function of the axial charge density}

Combining \eq(\ref{eq:WIAmom}) and (\ref{eq:WIPmom}), one finds 
\be\la{eq:kmuknu}
k_\mu k_\nu \int d^4x \; e^{ikx}\; \< A_\nu^b(0) A_\mu^a(x) \> = -m \delta^{ab}\<\bar\psi\psi\>
+4m^2 \int d^4x\;e^{ikx} \;\< P^b(0)P^a(x)\>.
\ee
Next we consider the difference of this relation at finite temperature
and at zero temperature.  The operator-product expansion indicates
that the most singular contributions arise from dimension four
operators.  By power counting, all the correlators appearing in
(\ref{eq:kmuknu}) are then expected to be finite.  More precisely, for
large $k_0$ and finite quark mass, all correlators in
(\ref{eq:kmuknu}) are of order $k_0^{-2}$.  The left-hand side of the
equation has a finite contribution when $k_0\to \infty$ given by the
$\int d^4x\, e^{ikx} \<A_0A_0\>|^T_0$ correlator, since it is multiplied by
$k_0^2$. On the right-hand side, the only surviving term is given by
the condensate. The coefficient of the ${\rm O}(k_0^{-2})$ term of the
$\int d^4x\,e^{ikx}\,\<A_0^a(0)A_0^b(x)\>|^T_0$ correlator must thus equal
$-m \delta^{ab}\<\bar\psi\psi\>|^T_0$ and it cannot contain logarithms
of $k_0$. To convert this statement into a property of the spectral
function, we use the spectral representation (see for instance \cite{Meyer:2011gj})
\be\la{eq:specrep}
\int d^4x \; e^{ikx}\; \< A_0^b(0) A_0^a(x) \>\Big|^T_0  = 
\delta^{ab}\int_{-\infty}^\infty d\omega\, \frac{\omega}{\omega^2 + k_0^2}\; \rho_{_{\rm A}}(\omega,k)\Big|^T_0.
\ee
The absence of logarithms in the coefficient of $k_0^{-2}$ on the left-hand side of \eq(\ref{eq:specrep})\footnote{This is confirmed by the two-loop calculation of
\cite{Chetyrkin:1985kn}. The most singular OPE term for the left-hand side of \eq(\ref{eq:specrep}) 
comes from the longitudinal channel, and is denoted 
 $\frac{k_0^2}{(k_0^2+\vec k^2)^2} C_2^{L} O_2^L$ in \cite{Chetyrkin:1985kn}.
In position space, the second derivative with respect to $x_0$ of this term is a contact term (plus terms 
of order $|x_0|$). It thus does not contribute to the spectral density at order $\omega^{-2}$.} indicates that 
 $\omega \rho_{_{\rm A}}(\omega,\vec k)|^T_0$ is integrable. Expanding the integrand on the right-hand side 
of \eq(\ref{eq:specrep}) to order $k_0^{-2}$ we then obtain
\be\la{sr:Aapdx}
\int_{-\infty}^\infty {d\omega\; \omega} \;\rho_{_{\rm A}}(\omega,\vec k)\Big|^{T}_{0} = -m\<\bar\psi\psi\>\Big|^{T}_{0}.
\ee

\section{Chiral perturbation theory predictions for finite-temperature observables\la{sec:apdxGL}}

The one-loop results of \cite{Gasser:1986vb} for the finite-temperature and finite-size effects on
the chiral observables can be written as
\begin{eqnarray}
&&\frac{{\cal O}(T,L)}{{\cal O}(0,\infty)} = 1-\nu_{\cal O}\frac{m_\pi^2}{f_\pi^2} \tilde g_1(m_\pi/T, m_\pi L),
\\
&&\nu_{f_\pi} = 1, \qquad \nu_{m_\pi} = -\frac{1}{4}, \qquad\nu_{\<\bar\psi\psi\>} = \frac{3}{2},
\\
&&\tilde g_1(x, y) = \frac{1}{(4\pi)^2}\sum_{n_1,n_2,n_3,n_4} \int^\infty_0 d\lambda \lambda^{-2} 
\exp{\left[-\lambda - \frac{1}{4\lambda}(y^2(n^2_1+n^2_2+n^2_3)+x^2n^2_4)\right]}.
\end{eqnarray}
On the right-hand side, $m_\pi$ and $f_\pi$ are understood to be the zero-temperature, infinite-volume quantities.
The sum runs over four integers, where the term $(n_1,n_2,n_3,n_4)=(0,0,0,0)$ is to be omitted. 
In addition to showing 
\be
\frac{f_\pi(T,\infty)}{f_\pi(0,\infty)},\qquad \quad \frac{m_\pi(T,\infty)}{m_\pi(0,\infty)}, \qquad \quad
\frac{\left<\bar{\psi}\psi\right>(T,\infty)}{\left<\bar{\psi}\psi\right>(0,\infty)},
\ee
as a function of $T$, we also display the curves 
\be
\frac{f_\pi(T,2/T)}{f_\pi(0,L_{\rm ref})} ,\qquad \quad \frac{m_\pi(T,2/T)}{m_\pi(0,L_{\rm ref})}, \qquad \quad 
\frac{\left<\bar{\psi}\psi\right>(T,2/T)}{\left<\bar{\psi}\psi\right>(0,L_{\rm ref})}
\ee
in \fig(\ref{fig:mfc}, \ref{fig:c2}), where $L_{\rm ref}$ corresponds to the spatial linear size of the $A_5$ ensemble.
In this way the finite size ($L=2/T$) of the spatial volume in our thermal ensembles
are taken into account in the comparison with the predictions of chiral perturbation theory.

\section{Tables}

\begin{table}[h]
\begin{center}
\begin{tabular}{c|c|c|c|c|c|c}
$6/g^2_0$ & $\kappa$ & $c_{\rm{sw}}$ & $a$ [fm] & $T$ [MeV] & $Z_A(g^2_0)$ & $\overline{m}^{\overline {\rm MS}}$ [MeV]\\ \hline\hline
5.20 & 0.13594 & 2.017147 & 0.0818(8) & 150(1) & 0.7703(57) & 15.4(4) \\
5.30 & 0.13636 & 1.909519 & 0.0693(6) & 177(2) & 0.7784(52) & 14.6(6) \\
5.355 & 0.13650 & 1.859618 & 0.0633(7) & 194(2) & 0.7826(49) & 14.7(6) \\
5.37 & 0.13652 & 1.846965 & 0.0618(7) & 199(2) & 0.7838(48) & 15.8(7) \\
5.38 & 0.13654 & 1.838739 & 0.0608(7) & 203(2) & 0.7845(48) & 15.5(9) \\
5.39 & 0.13656 & 1.830676 & 0.0599(6) & 206(2) & 0.7853(48) & 14.8(6) \\
5.40 & 0.13658 & 1.822771 & 0.0589(6) & 209(2) & 0.7860(47) & 16.8(7) \\
5.41 & 0.13660 & 1.815019 & 0.0580(6) & 213(2) & 0.7868(47) & 15.2(7) \\
5.42 & 0.13662 & 1.807416 & 0.0571(6) & 216(2) & 0.7875(46) & 14.0(7) \\
5.43 & 0.13664 & 1.799958 & 0.0562(6) & 219(2) & 0.7882(46) & 12.2(8) \\
5.44 & 0.13665 & 1.792642 & 0.0553(5) & 223(2) & 0.7889(45) & 14.2(10) \\
5.45 & 0.13666 & 1.785462 & 0.0544(5) & 226(2) & 0.7896(45) & 10.3(8) \\
5.47 & 0.13667 & 1.771499 & 0.0527(5) & 234(2) & 0.7910(44) & 15.4(9) \\ \hline\hline
\end{tabular}
\end{center}
\caption{Lattice parameters for the scan C1. All our finite-temperature lattices are 16$\times$$32^3$. The error on the lattice spacings and on the temperatures comes from interpolating a second order polynomial with the three known input values for $r_0/a$ evaluated at $6/g^2_0 = 5.20, 5.30, 5.50$ \cite{Fritzsch:2012wq}. The error shown on $\overline{m}^{\overline {\rm MS}}$ includes neither the uncertainty of the renormalization constants nor the error due to the scale setting. The latter two sources of error combine to be about  0.4-0.5 MeV in the whole range of $6/g^2_0$.}
\la{tab:C1}
\end{table}

\begin{table}[h]
\begin{center}
\begin{tabular}{c|c|c|c|c|c|c}
$6/g^2_0$ & $\kappa$ & $c_{\rm{sw}}$ & $a$ [fm] & $T$ [MeV] & $Z_A(g^2_0)$ & $\overline{m}^{\overline {\rm MS}}$ [MeV]\\ \hline\hline
5.30 & 0.13640 & 1.909519 & 0.0693(6) & 177(2) & 0.7784(52) & 8.2(8) \\
5.32 & 0.13646 & 1.890703 & 0.0671(7) & 183(2) & 0.7800(51) & 7.7(5) \\
5.33 & 0.13649 & 1.881590 & 0.0660(7) & 186(2) & 0.7808(50) & 5.2(10) \\
5.34 & 0.13651 & 1.872665 & 0.0649(7) & 189(2) & 0.7815(50) & 7.4(6) \\
5.35 & 0.13653 & 1.863922 & 0.0639(7) & 192(2) & 0.7823(49) & 7.9(6) \\
5.36 & 0.13655 & 1.855357 & 0.0628(7) & 195(2) & 0.7830(49) & 9.3(4) \\
5.37 & 0.13657 & 1.846965 & 0.0618(6) & 199(2) & 0.7838(48) & 9.7(9) \\
5.38 & 0.13659 & 1.838739 & 0.0608(7) & 203(2) & 0.7845(48) & 9.0(7) \\ \hline\hline
\end{tabular}
\end{center}
\caption{Lattice parameters with lower quark mass (scan D1; the lattice size is $16\times 32^3$ for each ensemble).  
The displayed errors have the same meaning as in Table \ref{tab:C1}.}
\la{tab:D1}
\end{table}

\begin{table}[h]
\centering
\begin{tabular}{l r}
\hline \hline
$6/g^2_0$ & 5.20\\
$\kappa$ & 0.13594 \\
$c_{\rm sw}$ & 2.017147 \\ \hline
$T$ [MeV] & 37.7(4) \\
$a$ [fm] & 0.0818(8) \\
$Z_A$ & 0.7703(57) \\
$\overline{m}^{\overline {\rm MS}} (\mu=2{\rm GeV})$ [MeV] & 14.7(3)\\ \hline
$m_{\pi}$ [MeV] & 305(5)\\
$f_{\pi}$ [MeV] & 93(2)\\
$\left|\left<\bar {\psi} \psi \right>^{\overline {\rm MS}}_{\rm GOR}\right|^{1/3}(\mu=2{\rm GeV})$ [MeV] & 364(7) \\ [.2cm] \hline
$\omega_{\bf 0}$ [MeV] & 294(4)\\
$f_{\pi, \bf{0}}$ [MeV] & 97(3)\\
$\left|\left<\bar {\psi} \psi \right>^{\overline {\rm MS}}_{\rm GOR, \bf 0}\right|^{1/3}(\mu=2{\rm GeV})$ [MeV] & 368(9) \\[.2cm] \hline
$u_f$ & 0.96(2) \\
$u_m$ & 0.92(6) \\
$u_f/u_m$ & 1.04(4) \\
$\omega_{\bf 0}/m_\pi$ & 0.96(2) \\ \hline \hline
\end{tabular}
\caption{Summary of results for the $64\times 32^3$ ensemble `A5'.}
\la{tab:A5pars}
\end{table}

\begin{table}[h]
\centering
\begin{tabular}{c|c|c|c|c|c|c|c|c}
$T$[MeV] & $G^{\overline {\rm MS}}_P(\beta/2)/T^3$ & $G_A(\beta/2)/T^3$ & $m_\pi(T)/T$ & $f_\pi(T)/T$ & $\left|\left<\bar {\psi} \psi \right>^{\overline {\rm MS}}_{\rm GOR}\right|^{1/3}/T$ & $u_f$ & $u_m$ & $u_f/u_m$\\ [.2cm] \hline \hline
150(1) & 26.5(7) & 0.38(2) & 2.15(4) & 0.59(2) & 2.40(5) & 0.88(2) & 0.84(2) & 1.04(3) \\
177(2) & 15.2(7) & 0.31(1) & 2.05(6) & 0.41(2) & 1.98(6) & 0.71(4) & 0.60(3) & 1.18(6)\\
194(2) & 8.1(4) & 0.35(1) & 2.4(1) & 0.26(2) & 1.67(7) & 0.43(3) & 0.32(3) & 1.3(1)\\
199(2) & 6.3(3) & 0.34(1) & 2.5(1) & 0.27(3) & 1.7(1) & 0.44(5) & 0.29(2) & 1.5(2)\\
203(2) & 4.5(2) & 0.394(8) & 3.13(8) & 0.15(2) & 1.4(1) & 0.24(3) & 0.18(2) & 1.3(2)\\
206(2) & 4.8(3) & 0.400(9) & 3.1(2) & 0.13(3) & 1.3(2) & 0.21(4) & 0.17(2) & 1.2(3)\\
209(2) & 5.8(4) & 0.38(1) & 2.4(3) & 0.18(3) & 1.3(2) & 0.28(4) & 0.28(4) & 1.0(2)\\
213(2) & 5.0(4) & 0.372(9) & 2.3(2) & 0.20(2) & 1.4(1) & 0.33(4) & 0.24(3) & 1.4(2)\\
216(2) & 4.3(3) & 0.384(8) & 3.4(1) & 0.12(3) & 1.3(2) & 0.18(5) & 0.14(1) & 1.3(2)\\
219(2) & 3.3(2) & 0.414(8) & 3.6(1) & 0.12(3) & 1.4(3) & 0.19(5) & 0.09(1) & 2.0(5)\\
223(2) & 4.0(3) & 0.401(9) & 2.6(2) & 0.21(3) & 1.6(2) & 0.32(4) & 0.16(3) & 2.0(3)\\
226(2) & 2.5(1) & 0.419(8) & 3.2(2) & 0.11(3) & 1.3(2) & 0.17(4) & 0.074(9) & 2.2(6)\\
234(2) & 2.8(2) & 0.429(7) & 3.1(3) & 0.13(1) & 1.3(1) & 0.19(2) & 0.11(2) & 1.6(3)\\ \hline \hline
\end{tabular}
\caption{Summary of numerical results for the  temperature scan C1. 
All errors given here are statistical and the uncertainty from the renormalization constants is not included.}
\la{tab:C1summ}
\end{table}

\begin{table}[h]
\centering
\begin{tabular}{c|c|c|c|c|c|c|c|c}
$T$[MeV] & $G^{\overline {\rm MS}}_P(\beta/2)/T^3$ & $G_A(\beta/2)/T^3$ & $m_\pi(T)/T$ & $f_\pi(T)/T$ & $\left|\left<\bar {\psi} \psi \right>^{\overline {\rm MS}}_{\rm GOR}\right|^{1/3}/T$ & $u_f$ & $u_m$ & $u_f/u_m$\\ [.2cm] \hline \hline
177(2) & 14.9(10) & 0.36(2) & 2.3(1) & 0.28(3) & 2.0(2) & 0.45(5) & 0.27(4) & 1.6(3)\\
183(2) & 11.5(9) & 0.35(1) & 2.26(9) & 0.28(4) & 2.0(2) & 0.46(7) & 0.22(2) & 2.0(3)\\
186(2) & 8.6(9) & 0.41(1) & 2.1(2) & 0.25(7) & 2.1(4) & 0.38(9) & 0.13(3) & 3(1)\\
189(2) & 9.9(9) & 0.37(1) & 2.38(9) & 0.26(2) & 2.1(1) & 0.42(3) & 0.18(2) & 2.3(3)\\
192(2) & 7.4(5) & 0.385(9) & 2.5(1) & 0.25(11) & 2.0(6) & 0.4(2) & 0.15(2) & 2.6(11)\\
195(2) & 8.9(5) & 0.369(8) & 2.5(1) & 0.20(2) & 1.6(1) & 0.33(4) & 0.21(2) & 1.6(2)\\
199(2) & 8.2(7) & 0.37(1) & 2.64(9) & 0.19(3) & 1.7(2) & 0.31(5) & 0.19(2) & 1.6(3)\\
203(2) & 7.1(5) & 0.39(1) & 2.7(1) & 0.19(3) & 1.8(2) & 0.30(5) & 0.15(2) & 2.0(4)\\  \hline \hline
\end{tabular}
\caption{Summary of numerical results for the $D_1$ temperature scan. 
All errors given here are statistical and the uncertainty from
the renormalization constants is not included.}
\la{tab:D1summ}
\end{table}

\clearpage 

\acknowledgments{ We are grateful for the access to the
  zero-temperature ensemble used here, made available to us through
  CLS.   We acknowledge the use of computing
  time for the generation of the gauge configurations on the JUGENE
  computer of the Gauss Centre for Supercomputing located at
  Forschungszentrum J\"ulich, Germany; the finite-temperature ensemble
  was generated within the John von Neumann Institute for Computing
  (NIC) project HMZ21. The correlation functions were computed on the
  dedicated QCD platform ``Wilson'' at the Institute for Nuclear
  Physics, University of Mainz. This work was supported by the
  \emph{Center for Computational Sciences in Mainz} as part of the
  Rhineland-Palatinate Research Initiative and by the DFG grant ME
  3622/2-1 \emph{Static and dynamic properties of QCD at finite
    temperature}.}

\bibliography{/Users/harvey/BIBLIO/viscobib.bib,../ANTHONY/mem.bib,../DANIEL/Version_2/bibliography_compendium.bib}

\begin{thebibliography}{54}
\expandafter\ifx\csname natexlab\endcsname\relax\def\natexlab#1{#1}\fi
\expandafter\ifx\csname bibnamefont\endcsname\relax
  \def\bibnamefont#1{#1}\fi
\expandafter\ifx\csname bibfnamefont\endcsname\relax
  \def\bibfnamefont#1{#1}\fi
\expandafter\ifx\csname citenamefont\endcsname\relax
  \def\citenamefont#1{#1}\fi
\expandafter\ifx\csname url\endcsname\relax
  \def\url#1{\texttt{#1}}\fi
\expandafter\ifx\csname urlprefix\endcsname\relax\def\urlprefix{URL }\fi
\providecommand{\bibinfo}[2]{#2}
\providecommand{\eprint}[2][]{\url{#2}}

\bibitem[{\citenamefont{Brambilla et~al.}(2014)\citenamefont{Brambilla,
  Eidelman, Foka, Gardner, Kronfeld et~al.}}]{Brambilla:2014aaa}
\bibinfo{author}{\bibfnamefont{N.}~\bibnamefont{Brambilla}},
  \bibinfo{author}{\bibfnamefont{S.}~\bibnamefont{Eidelman}},
  \bibinfo{author}{\bibfnamefont{P.}~\bibnamefont{Foka}},
  \bibinfo{author}{\bibfnamefont{S.}~\bibnamefont{Gardner}},
  \bibinfo{author}{\bibfnamefont{A.}~\bibnamefont{Kronfeld}},
  \bibnamefont{et~al.} (\bibinfo{year}{2014}), \eprint{1404.3723}.

\bibitem[{\citenamefont{Braun-Munzinger
  et~al.}(2012)\citenamefont{Braun-Munzinger, Friman, Karsch, Redlich, and
  Skokov}}]{BraunMunzinger:2011ta}
\bibinfo{author}{\bibfnamefont{P.}~\bibnamefont{Braun-Munzinger}},
  \bibinfo{author}{\bibfnamefont{B.}~\bibnamefont{Friman}},
  \bibinfo{author}{\bibfnamefont{F.}~\bibnamefont{Karsch}},
  \bibinfo{author}{\bibfnamefont{K.}~\bibnamefont{Redlich}}, \bibnamefont{and}
  \bibinfo{author}{\bibfnamefont{V.}~\bibnamefont{Skokov}},
  \bibinfo{journal}{Nucl.Phys.} \textbf{\bibinfo{volume}{A880}},
  \bibinfo{pages}{48} (\bibinfo{year}{2012}), \eprint{1111.5063}.

\bibitem[{\citenamefont{Stachel et~al.}(2013)\citenamefont{Stachel, Andronic,
  Braun-Munzinger, and Redlich}}]{Stachel:2013zma}
\bibinfo{author}{\bibfnamefont{J.}~\bibnamefont{Stachel}},
  \bibinfo{author}{\bibfnamefont{A.}~\bibnamefont{Andronic}},
  \bibinfo{author}{\bibfnamefont{P.}~\bibnamefont{Braun-Munzinger}},
  \bibnamefont{and} \bibinfo{author}{\bibfnamefont{K.}~\bibnamefont{Redlich}}
  (\bibinfo{year}{2013}), \eprint{1311.4662}.

\bibitem[{\citenamefont{Bazavov et~al.}(2012)}]{Bazavov:2012jq}
\bibinfo{author}{\bibfnamefont{A.}~\bibnamefont{Bazavov}} \bibnamefont{et~al.}
  (\bibinfo{collaboration}{HotQCD Collaboration}), \bibinfo{journal}{Phys.Rev.}
  \textbf{\bibinfo{volume}{D86}}, \bibinfo{pages}{034509}
  (\bibinfo{year}{2012}), \eprint{1203.0784}.

\bibitem[{\citenamefont{Borsanyi et~al.}(2014)\citenamefont{Borsanyi, Fodor,
  Hoelbling, Katz, Krieg et~al.}}]{Borsanyi:2013bia}
\bibinfo{author}{\bibfnamefont{S.}~\bibnamefont{Borsanyi}},
  \bibinfo{author}{\bibfnamefont{Z.}~\bibnamefont{Fodor}},
  \bibinfo{author}{\bibfnamefont{C.}~\bibnamefont{Hoelbling}},
  \bibinfo{author}{\bibfnamefont{S.~D.} \bibnamefont{Katz}},
  \bibinfo{author}{\bibfnamefont{S.}~\bibnamefont{Krieg}},
  \bibnamefont{et~al.}, \bibinfo{journal}{Phys.Lett.}
  \textbf{\bibinfo{volume}{B730}}, \bibinfo{pages}{99} (\bibinfo{year}{2014}),
  \eprint{1309.5258}.

\bibitem[{\citenamefont{Borsanyi et~al.}(2012)\citenamefont{Borsanyi, Fodor,
  Katz, Krieg, Ratti et~al.}}]{Borsanyi:2011sw}
\bibinfo{author}{\bibfnamefont{S.}~\bibnamefont{Borsanyi}},
  \bibinfo{author}{\bibfnamefont{Z.}~\bibnamefont{Fodor}},
  \bibinfo{author}{\bibfnamefont{S.~D.} \bibnamefont{Katz}},
  \bibinfo{author}{\bibfnamefont{S.}~\bibnamefont{Krieg}},
  \bibinfo{author}{\bibfnamefont{C.}~\bibnamefont{Ratti}},
  \bibnamefont{et~al.}, \bibinfo{journal}{JHEP}
  \textbf{\bibinfo{volume}{1201}}, \bibinfo{pages}{138} (\bibinfo{year}{2012}),
  \eprint{1112.4416}.

\bibitem[{\citenamefont{Meyer}(2011)}]{Meyer:2011gj}
\bibinfo{author}{\bibfnamefont{H.~B.} \bibnamefont{Meyer}},
  \bibinfo{journal}{Eur.Phys.J.} \textbf{\bibinfo{volume}{A47}},
  \bibinfo{pages}{86} (\bibinfo{year}{2011}), \eprint{1104.3708}.

\bibitem[{\citenamefont{Shuryak}(1990)}]{Shuryak:1990ie}
\bibinfo{author}{\bibfnamefont{E.~V.} \bibnamefont{Shuryak}},
  \bibinfo{journal}{Phys.Rev.} \textbf{\bibinfo{volume}{D42}},
  \bibinfo{pages}{1764} (\bibinfo{year}{1990}).

\bibitem[{\citenamefont{Goity and Leutwyler}(1989)}]{Goity:1989gs}
\bibinfo{author}{\bibfnamefont{J.}~\bibnamefont{Goity}} \bibnamefont{and}
  \bibinfo{author}{\bibfnamefont{H.}~\bibnamefont{Leutwyler}},
  \bibinfo{journal}{Phys.Lett.} \textbf{\bibinfo{volume}{B228}},
  \bibinfo{pages}{517} (\bibinfo{year}{1989}).

\bibitem[{\citenamefont{Schenk}(1991)}]{Schenk:1991xe}
\bibinfo{author}{\bibfnamefont{A.}~\bibnamefont{Schenk}},
  \bibinfo{journal}{Nucl.Phys.} \textbf{\bibinfo{volume}{B363}},
  \bibinfo{pages}{97} (\bibinfo{year}{1991}).

\bibitem[{\citenamefont{Schenk}(1993)}]{Schenk:1993ru}
\bibinfo{author}{\bibfnamefont{A.}~\bibnamefont{Schenk}},
  \bibinfo{journal}{Phys.Rev.} \textbf{\bibinfo{volume}{D47}},
  \bibinfo{pages}{5138} (\bibinfo{year}{1993}).

\bibitem[{\citenamefont{Toublan}(1997)}]{Toublan:1997rr}
\bibinfo{author}{\bibfnamefont{D.}~\bibnamefont{Toublan}},
  \bibinfo{journal}{Phys.Rev.} \textbf{\bibinfo{volume}{D56}},
  \bibinfo{pages}{5629} (\bibinfo{year}{1997}), \eprint{hep-ph/9706273}.

\bibitem[{\citenamefont{Pisarski and Tytgat}(1996)}]{Pisarski:1996mt}
\bibinfo{author}{\bibfnamefont{R.~D.} \bibnamefont{Pisarski}} \bibnamefont{and}
  \bibinfo{author}{\bibfnamefont{M.}~\bibnamefont{Tytgat}},
  \bibinfo{journal}{Phys.Rev.} \textbf{\bibinfo{volume}{D54}},
  \bibinfo{pages}{2989} (\bibinfo{year}{1996}), \eprint{hep-ph/9604404}.

\bibitem[{\citenamefont{Son and Stephanov}(2002{\natexlab{a}})}]{Son:2001ff}
\bibinfo{author}{\bibfnamefont{D.~T.} \bibnamefont{Son}} \bibnamefont{and}
  \bibinfo{author}{\bibfnamefont{M.~A.} \bibnamefont{Stephanov}},
  \bibinfo{journal}{Phys. Rev. Lett.} \textbf{\bibinfo{volume}{88}},
  \bibinfo{pages}{202302} (\bibinfo{year}{2002}{\natexlab{a}}),
  \eprint{hep-ph/0111100}.

\bibitem[{\citenamefont{Son and Stephanov}(2002{\natexlab{b}})}]{Son:2002ci}
\bibinfo{author}{\bibfnamefont{D.~T.} \bibnamefont{Son}} \bibnamefont{and}
  \bibinfo{author}{\bibfnamefont{M.~A.} \bibnamefont{Stephanov}},
  \bibinfo{journal}{Phys. Rev.} \textbf{\bibinfo{volume}{D66}},
  \bibinfo{pages}{076011} (\bibinfo{year}{2002}{\natexlab{b}}),
  \eprint{hep-ph/0204226}.

\bibitem[{\citenamefont{Brandt et~al.}(2013{\natexlab{a}})\citenamefont{Brandt,
  Francis, Meyer, Philipsen, and Wittig}}]{Brandt:2013mba}
\bibinfo{author}{\bibfnamefont{B.~B.} \bibnamefont{Brandt}},
  \bibinfo{author}{\bibfnamefont{A.}~\bibnamefont{Francis}},
  \bibinfo{author}{\bibfnamefont{H.~B.} \bibnamefont{Meyer}},
  \bibinfo{author}{\bibfnamefont{O.}~\bibnamefont{Philipsen}},
  \bibnamefont{and} \bibinfo{author}{\bibfnamefont{H.}~\bibnamefont{Wittig}}
  (\bibinfo{year}{2013}{\natexlab{a}}), \eprint{1310.8326}.

\bibitem[{\citenamefont{Burger et~al.}(2013)}]{Burger:2011zc}
\bibinfo{author}{\bibfnamefont{F.}~\bibnamefont{Burger}} \bibnamefont{et~al.}
  (\bibinfo{collaboration}{tmfT}), \bibinfo{journal}{Phys.Rev.}
  \textbf{\bibinfo{volume}{D87}}, \bibinfo{pages}{074508}
  (\bibinfo{year}{2013}), \eprint{1102.4530}.

\bibitem[{\citenamefont{Bornyakov et~al.}(2010)\citenamefont{Bornyakov,
  Horsley, Morozov, Nakamura, Polikarpov et~al.}}]{Bornyakov:2009qh}
\bibinfo{author}{\bibfnamefont{V.}~\bibnamefont{Bornyakov}},
  \bibinfo{author}{\bibfnamefont{R.}~\bibnamefont{Horsley}},
  \bibinfo{author}{\bibfnamefont{S.}~\bibnamefont{Morozov}},
  \bibinfo{author}{\bibfnamefont{Y.}~\bibnamefont{Nakamura}},
  \bibinfo{author}{\bibfnamefont{M.}~\bibnamefont{Polikarpov}},
  \bibnamefont{et~al.}, \bibinfo{journal}{Phys.Rev.}
  \textbf{\bibinfo{volume}{D82}}, \bibinfo{pages}{014504}
  (\bibinfo{year}{2010}), \eprint{0910.2392}.

\bibitem[{\citenamefont{Gasser and Leutwyler}(1987)}]{Gasser:1986vb}
\bibinfo{author}{\bibfnamefont{J.}~\bibnamefont{Gasser}} \bibnamefont{and}
  \bibinfo{author}{\bibfnamefont{H.}~\bibnamefont{Leutwyler}},
  \bibinfo{journal}{Phys.Lett.} \textbf{\bibinfo{volume}{B184}},
  \bibinfo{pages}{83} (\bibinfo{year}{1987}).

\bibitem[{\citenamefont{Luscher
  et~al.}(1997{\natexlab{a}})\citenamefont{L\"uscher, Sint, Sommer, and
  Wittig}}]{Luscher:1996jn}
\bibinfo{author}{\bibfnamefont{M.}~\bibnamefont{L\"uscher}},
  \bibinfo{author}{\bibfnamefont{S.}~\bibnamefont{Sint}},
  \bibinfo{author}{\bibfnamefont{R.}~\bibnamefont{Sommer}}, \bibnamefont{and}
  \bibinfo{author}{\bibfnamefont{H.}~\bibnamefont{Wittig}},
  \bibinfo{journal}{Nucl.Phys.} \textbf{\bibinfo{volume}{B491}},
  \bibinfo{pages}{344} (\bibinfo{year}{1997}{\natexlab{a}}),
  \eprint{hep-lat/9611015}.

\bibitem[{\citenamefont{Meyer}(2008)}]{Meyer:2008sn}
\bibinfo{author}{\bibfnamefont{H.~B.} \bibnamefont{Meyer}},
  \bibinfo{journal}{PoS} \textbf{\bibinfo{volume}{LAT08}}, \bibinfo{pages}{017}
  (\bibinfo{year}{2008}), \eprint{0809.5202}.

\bibitem[{\citenamefont{Bernecker and Meyer}(2011)}]{Bernecker:2011gh}
\bibinfo{author}{\bibfnamefont{D.}~\bibnamefont{Bernecker}} \bibnamefont{and}
  \bibinfo{author}{\bibfnamefont{H.~B.} \bibnamefont{Meyer}},
  \bibinfo{journal}{Eur.Phys.J.} \textbf{\bibinfo{volume}{A47}},
  \bibinfo{pages}{148} (\bibinfo{year}{2011}), \eprint{1107.4388}.

\bibitem[{\citenamefont{Brandt et~al.}(2013{\natexlab{b}})\citenamefont{Brandt,
  Francis, Meyer, and Wittig}}]{Brandt:2012jc}
\bibinfo{author}{\bibfnamefont{B.~B.} \bibnamefont{Brandt}},
  \bibinfo{author}{\bibfnamefont{A.}~\bibnamefont{Francis}},
  \bibinfo{author}{\bibfnamefont{H.~B.} \bibnamefont{Meyer}}, \bibnamefont{and}
  \bibinfo{author}{\bibfnamefont{H.}~\bibnamefont{Wittig}},
  \bibinfo{journal}{JHEP} \textbf{\bibinfo{volume}{1303}}, \bibinfo{pages}{100}
  (\bibinfo{year}{2013}{\natexlab{b}}), \eprint{1212.4200}.

\bibitem[{\citenamefont{Kapusta and Shuryak}(1994)}]{Kapusta:1993hq}
\bibinfo{author}{\bibfnamefont{J.~I.} \bibnamefont{Kapusta}} \bibnamefont{and}
  \bibinfo{author}{\bibfnamefont{E.~V.} \bibnamefont{Shuryak}},
  \bibinfo{journal}{Phys. Rev.} \textbf{\bibinfo{volume}{D49}},
  \bibinfo{pages}{4694} (\bibinfo{year}{1994}), \eprint{hep-ph/9312245}.

\bibitem[{\citenamefont{Jansen and Sommer}(1998)}]{Jansen:1998mx}
\bibinfo{author}{\bibfnamefont{K.}~\bibnamefont{Jansen}} \bibnamefont{and}
  \bibinfo{author}{\bibfnamefont{R.}~\bibnamefont{Sommer}}
  (\bibinfo{collaboration}{ALPHA collaboration}), \bibinfo{journal}{Nucl.Phys.}
  \textbf{\bibinfo{volume}{B530}}, \bibinfo{pages}{185} (\bibinfo{year}{1998}),
  \eprint{hep-lat/9803017}.

\bibitem[{\citenamefont{Hasenbusch}(2001)}]{Hasenbusch:2001ne}
\bibinfo{author}{\bibfnamefont{M.}~\bibnamefont{Hasenbusch}},
  \bibinfo{journal}{Phys.Lett.} \textbf{\bibinfo{volume}{B519}},
  \bibinfo{pages}{177} (\bibinfo{year}{2001}), \eprint{hep-lat/0107019}.

\bibitem[{\citenamefont{Hasenbusch and Jansen}(2003)}]{Hasenbusch:2002ai}
\bibinfo{author}{\bibfnamefont{M.}~\bibnamefont{Hasenbusch}} \bibnamefont{and}
  \bibinfo{author}{\bibfnamefont{K.}~\bibnamefont{Jansen}},
  \bibinfo{journal}{Nucl.Phys.} \textbf{\bibinfo{volume}{B659}},
  \bibinfo{pages}{299} (\bibinfo{year}{2003}), \eprint{hep-lat/0211042}.

\bibitem[{\citenamefont{Marinkovic and Schaefer}(2010)}]{Marinkovic:2010eg}
\bibinfo{author}{\bibfnamefont{M.}~\bibnamefont{Marinkovic}} \bibnamefont{and}
  \bibinfo{author}{\bibfnamefont{S.}~\bibnamefont{Schaefer}},
  \bibinfo{journal}{PoS} \textbf{\bibinfo{volume}{LATTICE2010}},
  \bibinfo{pages}{031} (\bibinfo{year}{2010}), \eprint{1011.0911}.

\bibitem[{CLS(2010)}]{CLScode}
\bibinfo{journal}{{http://luscher.web.cern.ch/luscher/DD-HMC/index.html}}
  (\bibinfo{year}{2010}).

\bibitem[{\citenamefont{Sommer}(1994)}]{Sommer:1993ce}
\bibinfo{author}{\bibfnamefont{R.}~\bibnamefont{Sommer}},
  \bibinfo{journal}{Nucl. Phys.} \textbf{\bibinfo{volume}{B411}},
  \bibinfo{pages}{839} (\bibinfo{year}{1994}), \eprint{hep-lat/9310022}.

\bibitem[{\citenamefont{Fritzsch et~al.}(2012)\citenamefont{Fritzsch, Knechtli,
  Leder, Marinkovic, Schaefer et~al.}}]{Fritzsch:2012wq}
\bibinfo{author}{\bibfnamefont{P.}~\bibnamefont{Fritzsch}},
  \bibinfo{author}{\bibfnamefont{F.}~\bibnamefont{Knechtli}},
  \bibinfo{author}{\bibfnamefont{B.}~\bibnamefont{Leder}},
  \bibinfo{author}{\bibfnamefont{M.}~\bibnamefont{Marinkovic}},
  \bibinfo{author}{\bibfnamefont{S.}~\bibnamefont{Schaefer}},
  \bibnamefont{et~al.}, \bibinfo{journal}{Nucl.Phys.}
  \textbf{\bibinfo{volume}{B865}}, \bibinfo{pages}{397} (\bibinfo{year}{2012}),
  \eprint{1205.5380}.

\bibitem[{\citenamefont{Bochicchio et~al.}(1985)\citenamefont{Bochicchio,
  Maiani, Martinelli, Rossi, and Testa}}]{Bochicchio:1985xa}
\bibinfo{author}{\bibfnamefont{M.}~\bibnamefont{Bochicchio}},
  \bibinfo{author}{\bibfnamefont{L.}~\bibnamefont{Maiani}},
  \bibinfo{author}{\bibfnamefont{G.}~\bibnamefont{Martinelli}},
  \bibinfo{author}{\bibfnamefont{G.~C.} \bibnamefont{Rossi}}, \bibnamefont{and}
  \bibinfo{author}{\bibfnamefont{M.}~\bibnamefont{Testa}},
  \bibinfo{journal}{Nucl. Phys.} \textbf{\bibinfo{volume}{B262}},
  \bibinfo{pages}{331} (\bibinfo{year}{1985}).

\bibitem[{\citenamefont{Luscher
  et~al.}(1997{\natexlab{b}})\citenamefont{Luscher, Sint, Sommer, Weisz, and
  Wolff}}]{Luscher:1996ug}
\bibinfo{author}{\bibfnamefont{M.}~\bibnamefont{L\"uscher}},
  \bibinfo{author}{\bibfnamefont{S.}~\bibnamefont{Sint}},
  \bibinfo{author}{\bibfnamefont{R.}~\bibnamefont{Sommer}},
  \bibinfo{author}{\bibfnamefont{P.}~\bibnamefont{Weisz}}, \bibnamefont{and}
  \bibinfo{author}{\bibfnamefont{U.}~\bibnamefont{Wolff}},
  \bibinfo{journal}{Nucl.Phys.} \textbf{\bibinfo{volume}{B491}},
  \bibinfo{pages}{323} (\bibinfo{year}{1997}{\natexlab{b}}),
  \eprint{hep-lat/9609035}.

\bibitem[{\citenamefont{Guagnelli et~al.}(2001)}]{Guagnelli:2000jw}
\bibinfo{author}{\bibfnamefont{M.}~\bibnamefont{Guagnelli}}
  \bibnamefont{et~al.} (\bibinfo{collaboration}{ALPHA Collaboration}),
  \bibinfo{journal}{Nucl.Phys.} \textbf{\bibinfo{volume}{B595}},
  \bibinfo{pages}{44} (\bibinfo{year}{2001}), \eprint{hep-lat/0009021}.

\bibitem[{\citenamefont{Della~Morte et~al.}(2005)\citenamefont{Della~Morte,
  Hoffmann, and Sommer}}]{DellaMorte:2005se}
\bibinfo{author}{\bibfnamefont{M.}~\bibnamefont{Della~Morte}},
  \bibinfo{author}{\bibfnamefont{R.}~\bibnamefont{Hoffmann}}, \bibnamefont{and}
  \bibinfo{author}{\bibfnamefont{R.}~\bibnamefont{Sommer}},
  \bibinfo{journal}{JHEP} \textbf{\bibinfo{volume}{0503}}, \bibinfo{pages}{029}
  (\bibinfo{year}{2005}), \eprint{hep-lat/0503003}.

\bibitem[{\citenamefont{Press et~al.}(1992)\citenamefont{Press, Flannery,
  Teukolsky, and Vetterling}}]{press_numerical_1992}
\bibinfo{author}{\bibfnamefont{W.}~\bibnamefont{Press}},
  \bibinfo{author}{\bibfnamefont{B.}~\bibnamefont{Flannery}},
  \bibinfo{author}{\bibfnamefont{S.}~\bibnamefont{Teukolsky}},
  \bibnamefont{and}
  \bibinfo{author}{\bibfnamefont{W.}~\bibnamefont{Vetterling}},
  \emph{\bibinfo{title}{Numerical Recipes in C: The Art of Scientific
  Computing}} (\bibinfo{publisher}{Cambridge University Press},
  \bibinfo{year}{1992}), ISBN \bibinfo{isbn}{0521431085},
  \urlprefix\url{http://www.amazon.ca/exec/obidos/redirect?tag=citeulike09-20&path=ASIN/0521431085}.

\bibitem[{\citenamefont{Gell-Mann et~al.}(1968)\citenamefont{Gell-Mann, Oakes,
  and Renner}}]{GellMann:1968rz}
\bibinfo{author}{\bibfnamefont{M.}~\bibnamefont{Gell-Mann}},
  \bibinfo{author}{\bibfnamefont{R.}~\bibnamefont{Oakes}}, \bibnamefont{and}
  \bibinfo{author}{\bibfnamefont{B.}~\bibnamefont{Renner}},
  \bibinfo{journal}{Phys.Rev.} \textbf{\bibinfo{volume}{175}},
  \bibinfo{pages}{2195} (\bibinfo{year}{1968}).

\bibitem[{\citenamefont{Gerber and Leutwyler}(1989)}]{Gerber:1988tt}
\bibinfo{author}{\bibfnamefont{P.}~\bibnamefont{Gerber}} \bibnamefont{and}
  \bibinfo{author}{\bibfnamefont{H.}~\bibnamefont{Leutwyler}},
  \bibinfo{journal}{Nucl.Phys.} \textbf{\bibinfo{volume}{B321}},
  \bibinfo{pages}{387} (\bibinfo{year}{1989}).

\bibitem[{\citenamefont{Asakawa et~al.}(2001)\citenamefont{Asakawa, Hatsuda,
  and Nakahara}}]{Asakawa:2000tr}
\bibinfo{author}{\bibfnamefont{M.}~\bibnamefont{Asakawa}},
  \bibinfo{author}{\bibfnamefont{T.}~\bibnamefont{Hatsuda}}, \bibnamefont{and}
  \bibinfo{author}{\bibfnamefont{Y.}~\bibnamefont{Nakahara}},
  \bibinfo{journal}{Prog. Part. Nucl. Phys.} \textbf{\bibinfo{volume}{46}},
  \bibinfo{pages}{459} (\bibinfo{year}{2001}), \eprint{hep-lat/0011040}.

\bibitem[{\citenamefont{Nakahara et~al.}(1999)\citenamefont{Nakahara, Asakawa,
  and Hatsuda}}]{Nakahara:1999vy}
\bibinfo{author}{\bibfnamefont{Y.}~\bibnamefont{Nakahara}},
  \bibinfo{author}{\bibfnamefont{M.}~\bibnamefont{Asakawa}}, \bibnamefont{and}
  \bibinfo{author}{\bibfnamefont{T.}~\bibnamefont{Hatsuda}},
  \bibinfo{journal}{Phys. Rev.} \textbf{\bibinfo{volume}{D60}},
  \bibinfo{pages}{091503} (\bibinfo{year}{1999}), \eprint{hep-lat/9905034}.

\bibitem[{\citenamefont{Yamazaki et~al.}(2002)}]{Yamazaki:2001er}
\bibinfo{author}{\bibfnamefont{T.}~\bibnamefont{Yamazaki}} \bibnamefont{et~al.}
  (\bibinfo{collaboration}{CP-PACS Collaboration}),
  \bibinfo{journal}{Phys.Rev.} \textbf{\bibinfo{volume}{D65}},
  \bibinfo{pages}{014501} (\bibinfo{year}{2002}), \eprint{hep-lat/0105030}.

\bibitem[{\citenamefont{Fiebig}(2002)}]{Fiebig:2002sp}
\bibinfo{author}{\bibfnamefont{H.~R.} \bibnamefont{Fiebig}},
  \bibinfo{journal}{Phys.Rev.} \textbf{\bibinfo{volume}{D65}},
  \bibinfo{pages}{094512} (\bibinfo{year}{2002}), \eprint{hep-lat/0204004}.

\bibitem[{\citenamefont{Sasaki et~al.}(2005)\citenamefont{Sasaki, Sasaki, and
  Hatsuda}}]{Sasaki:2005ap}
\bibinfo{author}{\bibfnamefont{K.}~\bibnamefont{Sasaki}},
  \bibinfo{author}{\bibfnamefont{S.}~\bibnamefont{Sasaki}}, \bibnamefont{and}
  \bibinfo{author}{\bibfnamefont{T.}~\bibnamefont{Hatsuda}},
  \bibinfo{journal}{Phys.Lett.} \textbf{\bibinfo{volume}{B623}},
  \bibinfo{pages}{208} (\bibinfo{year}{2005}), \eprint{hep-lat/0504020}.

\bibitem[{\citenamefont{Karsch et~al.}(2002)\citenamefont{Karsch, Laermann,
  Petreczky, Stickan, and Wetzorke}}]{Karsch:2001uw}
\bibinfo{author}{\bibfnamefont{F.}~\bibnamefont{Karsch}},
  \bibinfo{author}{\bibfnamefont{E.}~\bibnamefont{Laermann}},
  \bibinfo{author}{\bibfnamefont{P.}~\bibnamefont{Petreczky}},
  \bibinfo{author}{\bibfnamefont{S.}~\bibnamefont{Stickan}}, \bibnamefont{and}
  \bibinfo{author}{\bibfnamefont{I.}~\bibnamefont{Wetzorke}},
  \bibinfo{journal}{Phys.Lett.} \textbf{\bibinfo{volume}{B530}},
  \bibinfo{pages}{147} (\bibinfo{year}{2002}), \eprint{hep-lat/0110208}.

\bibitem[{\citenamefont{Aarts et~al.}(2007{\natexlab{a}})\citenamefont{Aarts,
  Allton, Foley, Hands, and Kim}}]{Aarts:2007wj}
\bibinfo{author}{\bibfnamefont{G.}~\bibnamefont{Aarts}},
  \bibinfo{author}{\bibfnamefont{C.}~\bibnamefont{Allton}},
  \bibinfo{author}{\bibfnamefont{J.}~\bibnamefont{Foley}},
  \bibinfo{author}{\bibfnamefont{S.}~\bibnamefont{Hands}}, \bibnamefont{and}
  \bibinfo{author}{\bibfnamefont{S.}~\bibnamefont{Kim}},
  \bibinfo{journal}{Phys. Rev. Lett.} \textbf{\bibinfo{volume}{99}},
  \bibinfo{pages}{022002} (\bibinfo{year}{2007}{\natexlab{a}}),
  \eprint{hep-lat/0703008}.

\bibitem[{\citenamefont{Aarts et~al.}(2007{\natexlab{b}})\citenamefont{Aarts,
  Allton, Oktay, Peardon, and Skullerud}}]{Aarts:2007pk}
\bibinfo{author}{\bibfnamefont{G.}~\bibnamefont{Aarts}},
  \bibinfo{author}{\bibfnamefont{C.}~\bibnamefont{Allton}},
  \bibinfo{author}{\bibfnamefont{M.~B.} \bibnamefont{Oktay}},
  \bibinfo{author}{\bibfnamefont{M.}~\bibnamefont{Peardon}}, \bibnamefont{and}
  \bibinfo{author}{\bibfnamefont{J.-I.} \bibnamefont{Skullerud}},
  \bibinfo{journal}{Phys. Rev.} \textbf{\bibinfo{volume}{D76}},
  \bibinfo{pages}{094513} (\bibinfo{year}{2007}{\natexlab{b}}),
  \eprint{0705.2198}.

\bibitem[{\citenamefont{Ding et~al.}(2011{\natexlab{a}})\citenamefont{Ding,
  Francis, Kaczmarek, Karsch, Laermann et~al.}}]{Ding:2010ga}
\bibinfo{author}{\bibfnamefont{H.-T.} \bibnamefont{Ding}},
  \bibinfo{author}{\bibfnamefont{A.}~\bibnamefont{Francis}},
  \bibinfo{author}{\bibfnamefont{O.}~\bibnamefont{Kaczmarek}},
  \bibinfo{author}{\bibfnamefont{F.}~\bibnamefont{Karsch}},
  \bibinfo{author}{\bibfnamefont{E.}~\bibnamefont{Laermann}},
  \bibnamefont{et~al.}, \bibinfo{journal}{Phys.Rev.}
  \textbf{\bibinfo{volume}{D83}}, \bibinfo{pages}{034504}
  (\bibinfo{year}{2011}{\natexlab{a}}), \eprint{1012.4963}.

\bibitem[{\citenamefont{Ding et~al.}(2011{\natexlab{b}})\citenamefont{Ding,
  Francis, Kaczmarek, Karsch, Satz et~al.}}]{Ding:2011hr}
\bibinfo{author}{\bibfnamefont{H.}~\bibnamefont{Ding}},
  \bibinfo{author}{\bibfnamefont{A.}~\bibnamefont{Francis}},
  \bibinfo{author}{\bibfnamefont{O.}~\bibnamefont{Kaczmarek}},
  \bibinfo{author}{\bibfnamefont{F.}~\bibnamefont{Karsch}},
  \bibinfo{author}{\bibfnamefont{H.}~\bibnamefont{Satz}}, \bibnamefont{et~al.},
  \bibinfo{journal}{J.Phys.} \textbf{\bibinfo{volume}{G38}},
  \bibinfo{pages}{124070} (\bibinfo{year}{2011}{\natexlab{b}}),
  \eprint{1107.0311}.

\bibitem[{\citenamefont{Ding et~al.}(2012)\citenamefont{Ding, Francis,
  Kaczmarek, Karsch, Satz et~al.}}]{Ding:2012sp}
\bibinfo{author}{\bibfnamefont{H.}~\bibnamefont{Ding}},
  \bibinfo{author}{\bibfnamefont{A.}~\bibnamefont{Francis}},
  \bibinfo{author}{\bibfnamefont{O.}~\bibnamefont{Kaczmarek}},
  \bibinfo{author}{\bibfnamefont{F.}~\bibnamefont{Karsch}},
  \bibinfo{author}{\bibfnamefont{H.}~\bibnamefont{Satz}}, \bibnamefont{et~al.},
  \bibinfo{journal}{Phys.Rev.} \textbf{\bibinfo{volume}{D86}},
  \bibinfo{pages}{014509} (\bibinfo{year}{2012}), \eprint{1204.4945}.

\bibitem[{\citenamefont{Karsch et~al.}(2003)\citenamefont{Karsch, Laermann,
  Petreczky, and Stickan}}]{Karsch:2003wy}
\bibinfo{author}{\bibfnamefont{F.}~\bibnamefont{Karsch}},
  \bibinfo{author}{\bibfnamefont{E.}~\bibnamefont{Laermann}},
  \bibinfo{author}{\bibfnamefont{P.}~\bibnamefont{Petreczky}},
  \bibnamefont{and} \bibinfo{author}{\bibfnamefont{S.}~\bibnamefont{Stickan}},
  \bibinfo{journal}{Phys.Rev.} \textbf{\bibinfo{volume}{D68}},
  \bibinfo{pages}{014504} (\bibinfo{year}{2003}), \eprint{hep-lat/0303017}.

\bibitem[{\citenamefont{Aarts and Martinez~Resco}(2005)}]{Aarts:2005hg}
\bibinfo{author}{\bibfnamefont{G.}~\bibnamefont{Aarts}} \bibnamefont{and}
  \bibinfo{author}{\bibfnamefont{J.~M.} \bibnamefont{Martinez~Resco}},
  \bibinfo{journal}{Nucl. Phys.} \textbf{\bibinfo{volume}{B726}},
  \bibinfo{pages}{93} (\bibinfo{year}{2005}), \eprint{hep-lat/0507004}.

\bibitem[{\citenamefont{Bhattacharya et~al.}(2014)\citenamefont{Bhattacharya,
  Buchoff, Christ, Ding, Gupta et~al.}}]{Bhattacharya:2014ara}
\bibinfo{author}{\bibfnamefont{T.}~\bibnamefont{Bhattacharya}},
  \bibinfo{author}{\bibfnamefont{M.~I.} \bibnamefont{Buchoff}},
  \bibinfo{author}{\bibfnamefont{N.~H.} \bibnamefont{Christ}},
  \bibinfo{author}{\bibfnamefont{H.~T.} \bibnamefont{Ding}},
  \bibinfo{author}{\bibfnamefont{R.}~\bibnamefont{Gupta}}, \bibnamefont{et~al.}
  (\bibinfo{year}{2014}), \eprint{1402.5175}.

\bibitem[{\citenamefont{Luscher}(1998)}]{Luscher:1998pe}
\bibinfo{author}{\bibfnamefont{M.}~\bibnamefont{L\"uscher}}
  (\bibinfo{year}{1998}), \eprint{hep-lat/9802029}.

\bibitem[{\citenamefont{Chetyrkin et~al.}(1985)\citenamefont{Chetyrkin,
  Spiridonov, and Gorishnii}}]{Chetyrkin:1985kn}
\bibinfo{author}{\bibfnamefont{K.}~\bibnamefont{Chetyrkin}},
  \bibinfo{author}{\bibfnamefont{V.}~\bibnamefont{Spiridonov}},
  \bibnamefont{and}
  \bibinfo{author}{\bibfnamefont{S.}~\bibnamefont{Gorishnii}},
  \bibinfo{journal}{Phys.Lett.} \textbf{\bibinfo{volume}{B160}},
  \bibinfo{pages}{149} (\bibinfo{year}{1985}).

\end{thebibliography}

\end{document}